\documentclass[acmsmall,screen]{acmart}

\AtBeginDocument{%
  \providecommand\BibTeX{{%
    \normalfont B\kern-0.5em{\scshape i\kern-0.25em b}\kern-0.8em\TeX}}}

\setcopyright{acmcopyright}
\copyrightyear{2025}
\acmYear{2025}
\acmDOI{10.114/1122445.1122456}

\acmJournal{JACM}
\acmVolume{37}
\acmNumber{4}
\acmMonth{7}

\usepackage{color}
\usepackage{multirow}
\usepackage{threeparttable} 
\usepackage{subfigure}
\usepackage{algorithmic}
\usepackage{graphicx}
\usepackage{textcomp}
\usepackage{xcolor}
\usepackage{stfloats}
\usepackage{bbding}
\usepackage{pifont}
\usepackage{hyperref}
\usepackage{makecell}
\usepackage{booktabs}
\usepackage{tabularx}
\usepackage{amsmath}
\usepackage[linesnumbered,ruled]{algorithm2e}
\usepackage{svg}
\usepackage{url}
\usepackage{float}
\usepackage{xspace}
\usepackage{tcolorbox}
\usepackage{tikz}
\usepackage{wrapfig}

\newcommand*{\circled}[1]{\lower.7ex\hbox{\tikz\draw (0pt, 0pt)%
    circle (.45em) node {\makebox[0.6em][c]{\small #1}};}}

\newcommand{\cxmark}{\ding{51}\textsuperscript{\kern-0.55em\ding{55}}}
\definecolor{dgreen}{RGB}{46,175,87}
\definecolor{customred}{HTML}{db5a6b}
\definecolor{customblue}{HTML}{1685a9}

\begin{document}

\title{Beyond the Edge of Function: Unraveling the Patterns of Type Recovery in Binary Code}


\author{Gangyang~Li}
\authornote{Both authors contributed equally to this research.}
\email{ligangyang@mail.ustc.edu.cn}
\affiliation{%
    \institution{University of Science and Technology of China}
    \city{Hefei}
    \country{Anhui, China}
}
\author{Xiuwei~Shang}
\authornotemark[1]
\email{shangxw@mail.ustc.edu.cn, xwshang@smu.edu.sg}
\affiliation{%
  \institution{University of Science and Technology of China}
  \country{Hefei, China}
  }
\affiliation{
  \institution{Singapore Management University}
  \country{Singapore}
}
\author{Shaoyin~Cheng}
\authornote{Corresponding author.}
\email{sycheng@ustc.edu.cn}
\affiliation{%
  \institution{University of Science and Technology of China, Anhui Province Key Laboratory of Digital Security}
    \city{Hefei}
    \country{Anhui, China}
}
\author{Junqi~Zhang}
\authornotemark[2]
\email{jqzh@ustc.edu.cn}
\affiliation{%
  \institution{University of Science and Technology of China, Anhui Province Key Laboratory of Digital Security}
    \city{Hefei}
    \country{Anhui, China}
}
\author{Li~Hu}
\email{pdxbshx@mail.ustc.edu.cn}
\affiliation{%
  \institution{University of Science and Technology of China}
    \city{Hefei}
    \country{Anhui, China}
}
\author{Xu~Zhu}
\email{zhuxu24@mail.ustc.edu.cn}
\affiliation{%
  \institution{University of Science and Technology of China}
    \city{Hefei}
    \country{Anhui, China}
}
\author{Weiming~Zhang}
\email{zhangwm@ustc.edu.cn}
\affiliation{%
  \institution{University of Science and Technology of China, Anhui Province Key Laboratory of Digital Security}
    \city{Hefei}
    \country{Anhui, China}
}
\author{Nenghai~Yu}
\email{ynh@ustc.edu.cn}
\affiliation{%
  \institution{University of Science and Technology of China, Anhui Province Key Laboratory of Digital Security}
    \city{Hefei}
    \country{Anhui, China}
}

\renewcommand{\shortauthors}{Gangyang Li et al.}


\begin{abstract}

Type recovery is a crucial step in binary code analysis, holding significant importance for reverse engineering and various security applications. Existing works typically simply target type identifiers within binary code and achieve type recovery by analyzing variable characteristics within functions. However, we find that the types in real-world binary programs are more complex and often follow specific distribution patterns.

In this paper, to gain a profound understanding of the variable type recovery problem in binary code, we first conduct a comprehensive empirical study. We utilize the TYDA dataset, which includes 163,643 binary programs across four architectures and four compiler optimization options, fully reflecting the complexity and diversity of real-world programs. We carefully study the unique patterns that characterize types and variables in binary code, and also investigate the impact of compiler optimizations, yielding many valuable insights.
Based on our empirical findings, we propose \textbf{ByteTR}, a framework for recovering variable types in binary code. We {map the infinite set of target types to a finite, predictable set to obtain a balanced type dataset} and perform static program analysis to tackle the impact of compiler optimizations on variable storage. In light of the ubiquity of variable propagation across functions observed in our study, ByteTR conducts inter-procedural analysis to trace variable propagation and employs a gated graph neural network to capture long-range data flow dependencies for variable type recovery. We conduct extensive experiments to evaluate the performance of ByteTR. The results demonstrate that ByteTR outperforms state-of-the-art works in both effectiveness and efficiency. Moreover, in {real-world} {capture-the-flag (CTF)} {vulnerability analysis} case, the pseudo code optimized by ByteTR significantly improves readability, surpassing leading tools IDA Pro and Ghidra {as well as methods DIRTY and StateFormer}, {demonstrating the substantial practical applicability}.

\end{abstract}


\maketitle

\section{Introduction \label{sec:introduction}}

Type recovery is a fundamental and crucial step in the decompilation process \cite{liu2020far}, providing high-level semantic abstractions and complex expression constraints of the target entity. It plays a pivotal role in various downstream security tasks, including vulnerability discovery \cite{vadayath2022arbiter, luo2023vulhawk}, malware detection \cite{avllazagaj2021malware, garcia2018lightweight, cao2020benign}, and taint analysis \cite{sang2024airtaint, wang2023taintmini, liang2022pata} in binary code. 
In reverse engineering scenarios, we typically need to analyze closed-source binary programs, which present multiple significant challenges. The obstacles include the inability to access high-level abstractions from source code, the distortion of code structure resulting from compiler optimizations, and the removal of symbol information from binary files. These compounding factors make type recovery in binary code both a crucial and technically challenging task.

Traditional type recovery approaches \cite{IDA_PRO,GHIDRA,BIN_JA,srinivasan2014recovery} primarily rely on manually crafted heuristic rules, deriving variable type information through dynamic or static analysis of program behavioral characteristics. However, when confronting diverse Instruction Set Architectures (ISA) and Application Binary Interfaces (ABI), the complexity of these rules grows exponentially. Furthermore, various compiler options subtly influence the program behavior in binary code, causing rule-based methods to frequently produce ambiguous or uncertain type inference results.

In recent years, with the flourishing of generative large language models, numerous deep learning-based solutions for type recovery have emerged in the community. DIRTY \cite{chen2021augmenting} employs a transformer-based model to recover variable names and type information, including primitive types and user-defined composite types, from pseudo code. {TYGR \cite{zhu2024tygr} explores variable memory access patterns using symbolic execution, extracting a one-hot encoding of 16 fixed features as nodes to build a function-level data flow graph to describe program behavior for type inference.} StateFormer \cite{pei2021stateformer} introduces a pre-training task based on Generative State Modeling to learn binary code semantics and infer type information through model fine-tuning.
{Debin \cite{he2018debin} uses a Conditional Random Field to predict and recover variable attributes from stripped binaries. OSPREY \cite{zhang2021osprey}, ReSym \cite{xie2024resym}, and llasm \cite{sha2024llasm} are also representative works that explored using probabilistic graphical models, large pre-trained models, and encoder-decoder models for type recovery and binary program analysis, respectively.}

These approaches leverage deep learning techniques by employing feature engineering to extract type information from binary files and classifying each discovered type as a distinct label. However, they fail to consider the inherent characteristics of types, such as their natural distribution in binary code, recursive definitions and alias attributes, and various compiler optimizations that affect variable type information. All these factors significantly impact the effectiveness of type recovery. Indeed, DIRTY demonstrates limited generalization beyond its training dataset, {TYGR only supports processing variables stored in stack memory,} while Debin and StateFormer do not perform well on real-world binary programs. This situation highlights a significant {gap} in existing approaches: the lack of a comprehensive assessment of the intrinsic features of type information naturally embedded in binary files. Such an assessment is crucial for \textbf{Unraveling the patterns of type recovery in binary code} and devising effective solutions.

In this paper, we first conduct a comprehensive empirical analysis from three perspectives: types, variables, and compilation, to investigate their intrinsic properties and relationships. Our analysis reveals that types in binary files exhibit naturally unbalanced distribution in quantitative characteristics while following specific distribution patterns. Structure types exhibit Memory Overlap in layout, and Locality of Reference when accessing their members, making some structure variables indistinguishable. Moreover, we discover that nearly half of the variables propagate across multiple functions. Finally, while compilation optimization improves program execution efficiency, it affects variable storage patterns, shifting from stack-based storage to register-based storage. 

Based on our empirical findings, we reformulate the variable type recovery problem in binary code as a type prediction task and propose \textbf{ByteTR} (\textbf{\underbar{B}}e\textbf{\underbar{y}}ond \textbf{\underbar{t}}he \textbf{\underbar{e}}dge of function for \textbf{\underbar{T}}ype \textbf{\underbar{R}}ecovery), a novel deep learning-based framework for type recovery in binary code. In ByteTR, to address the challenge of imbalanced type distribution, we decompose the custom types and alias types into primitive types as prediction targets (e.g., \texttt{FILE *} $\rightarrow$ \texttt{struct *}, \texttt{ssize\_t} $\rightarrow$ \texttt{long int}). 

In response to the observation of widespread variable cross-function propagation, we aim to track variables across functions. {However, program analysis at the level of IR lifted from binary code presents significant challenges. Unlike source code, which features explicit variable definitions, binary analysis must operate on low-level entities such as memory units and registers. Furthermore, binary code lacks explicit function call abstractions, while instruction semantics are fine-grained and often entail implicit side effects. To address these issues, we propose the BytePA algorithm. This approach uses variable expressions to provide a unified model for binary-level variables and reconstructs function call details. Consequently, we are able to track variable behaviors beyond function edges to construct an inter-procedural Variable Propagation Graph.}

To account for compiler optimization effects on variable storage patterns, {we use a single variable expression to uniformly model both stack-based and register-based variables.} In addition, we employ Gated Graph Neural Networks (GGNN) \cite{li2015gated} to capture detailed variable behavioral features, thus facilitating the inference of variable types. {All these design choices are supported by data, which enables our method to effectively handle real-world binary programs.} {Figure \ref{fig:realworldcase} showcases the type recovery results of ByteTR on a real-world vulnerability analysis case, further demonstrating that it surpasses previous work and validating the effectiveness of our design.} 

We evaluate ByteTR on the TYDA \cite{zhu2024tygr} dataset, comprising 4 architectures and 4 optimization options, with more than 21 million functions and 105 million variables. 
{The TYDA dataset} eliminated the high function duplication rate found in previous datasets {\cite{pal2024len, lacomis2019dire, chen2021augmenting}}, and has been peer-reviewed {in the field of binary code analysis}. {Experimental} results show that ByteTR can recover variable type with an average precision of 75.84\%, outperforming baseline DIRTY, StateFormer, and TYGR by 32.63\%, 11.93\%, and 23.29\%, {as well as DeepSeek-v3 and GPT-4o by 15.26\% and 13.96\%, and Ghidra and IDA Pro by 26.82\%  and 23.62\%} respectively. In addition, we apply ByteTR to CTF challenges and find that pseudo code optimized by ByteTR significantly improves readability.

Our contributions can be summarized as follows:
\begin{itemize} 
    \item \textbf{Empirical Analysis}. We present the first comprehensive empirical study on variable type in binary code. Our findings not only rectify several misconceptions but also provide valuable insights into leveraging deep learning techniques for type recovery task.

    \item \textbf{Innovative Algorithm}. Based on our findings from empirical analysis, we propose a novel algorithm, BytePA, to construct the Variable Propagation Graph (VPG). This algorithm employs an inter-procedural analytical method capable of simultaneously processing variable definition patterns for both stack-based and register-based storage, while systematically tracking their propagation paths between functions.

    \item \textbf{Effective Framework}. We present ByteTR, an innovative binary code variable type recovery framework. It transforms VPG into Variable Semantic Graph (VSG) to model variable features, and utilizes GGNN to achieve variable-level semantic representation. This approach clearly represents the operational patterns and data flow characteristics of variable and significantly improves the precision of variable type prediction.
    
    \item \textbf{Comprehensive Experiment}. 
    We conduct comprehensive experiments across different architectures and optimization levels to evaluate the performance of ByteTR. The results show that ByteTR outperforms the baselines and achieves state-of-the-art in terms of efficiency and effectiveness. In the real-world case, pseudo code optimized with ByteTR improves readability over IDA Pro, Ghidra, DIRTY, and StateFormer.

\end{itemize}

\noindent\textbf{Paper Organization.} The rest of this paper is organized as follows: 
In Section \ref{sec:background}, we elaborate on the research background, theoretical foundations, and primary motivations driving this study. Section \ref{sec:empiricalanalysis} systematically presents our empirical study. Section \ref{sec:methodology} focuses on the technical details of ByteTR, providing its core design principles and implementation specifics. In Section \ref{sec:evaluation}, we conduct a rigorous evaluation of ByteTR's performance. Section \ref{sec:disandlimi} discusses the scalability and generality of our method, and analyzes our limitations and directions for future work. Finally, Section \ref{sec:relatedworks} reviews related work, and Section \ref{sec:conclusion} summarizes the main contributions and findings of this study.
 
\noindent\textbf{Artifact Availability.} The implementation of ByteTR and all experimental scripts are publicly available at our GitHub repository\footnote{\url{https://github.com/giles-one/ByteTR}} to facilitate reproducibility and future research in this domain.

\section{Background \label{sec:background}}

Before diving into the technical details of our work, this section presents the necessary background knowledge encompassing Decompilation, Type System, and Binary Code Semantics Representation.

\subsection{Decompilation \label{sec:decompilation}}

Decompilation is a technical process of transforming machine code or bytecode from compiled programs back into high-level programming language representation. Although both decompilers and disassemblers \cite{OBJDUMP} can generate human-readable output, decompilers are capable of producing higher-level abstract representations, resulting in more concise and comprehensible code structures.

Modern decompilation techniques have made significant advances. Both commercial solutions, such as HexRays IDA Pro \cite{IDA_PRO} and Binary Ninja \cite{BIN_JA}, as well as open-source tool Ghidra \cite{GHIDRA}, can reconstruct binary code into C-style pseudo code. However, current decompilers still exhibit limitations in type inference for variables. For instance, in HexRays IDA, the leading decompiler in the industry, 64-bit pointers and structures are frequently misidentified as the \texttt{\_\_int64} integer type, often necessitating interactive human analysis and correction.

Beyond traditional heuristics-based decompilation pipelines, prior work has also begun exploring machine learning–based paradigms across various sub-tasks. {This work focuses on the problem of variable type inference in decompilers and explores using deep learning models to recover the original types of variables. We formulate type recovery as a classification task by predefining a list of target types, allowing the model to learn the behavioral semantics of variables to infer their corresponding type from this list. When applying our framework, we first identify target variables using location expressions obtained from the decompiler, and then feed the predicted types back to assist in generating pseudo code with accurate type annotations.}

\subsection{Type System \label{sec:typesystem}}

A type system serves as a formal logical framework that assigns specific properties, known as types (such as integers, floating-point numbers, and strings), to various language entities through a set of well-defined rules. In the realm of real-world software development, carefully designed type systems have become indispensable, as they address programmers' fundamental need to abstract and constrain computational expressions. The identification of variable types within binary code plays a crucial role in semantic understanding, significantly influencing various downstream analysis tasks. Given that contemporary decompilers predominantly transform machine code into C-style pseudo code, our analysis centers on the C language type system and its characteristics. 

\textbf{Primitive Types}. Primitive types are standard types embedded within programming languages, with predefined type identifiers. Primitive types are almost invariably value types. In the C programming language, there are four fundamental data types: \texttt{char}, \texttt{int}, \texttt{float}, and \texttt{double}, {along with} four type modifiers: \texttt{signed}/\texttt{unsigned} {for extending integer sign representation}, and \texttt{short}/\texttt{long} {for adjusting integer bit-width}. {The \texttt{char} type is a special integer type typically fixed at 8 bits, primarily used for character storage. While it can be modified with sign specifiers, its bit-width cannot be extended. The C standard only mandates that the \texttt{int} type must have a minimum bit-width of 16 bits. Due to platform dependencies, compiler implementations, and historical reasons, int defaults to 16 bits on 16-bit systems and 32 bits on 32-bit and 64-bit systems. The \texttt{int} type supports both sign modification and bit-width extension, supporting all four modifiers. Unlike integers, the sign representation and precision of \texttt{float} and \texttt{double} are governed by the IEEE 754 \cite{IEEE754} standard and generally do not support type modifiers for extension.} This design paradigm has influenced numerous subsequent programming languages.

\textbf{Composite Types}. Composite types represent advanced type entities derived from the composition of primitive types. While this compositional approach theoretically enables the generation of infinite type variations, these types typically require human-assigned unique identifiers or names for distinction, as exemplified by \texttt{struct} and \texttt{enum} in C language. Furthermore, array types and pointer types, which generate new types through recursive definitions, are also composite types.

\textbf{Alias Types}. Type aliases, implemented through the \texttt{typedef} keyword, do not create new types but rather establish references to existing types, maintaining complete type equivalence. For instance, the declaration "\texttt{typedef long int time\_t}" creates an alias commonly used to represent time variables, while fundamentally retaining its underlying \texttt{long int} type. While such abstraction facilitates programming by improving code readability and maintenance, it actually introduces challenges in type prediction in reverse engineering. During the lexical analysis phase of compilation, type aliases are systematically replaced with their original types for further processing.

Types make restrictions on the behavioral characteristics of the variables they are assigned to, which is the fundamental theory for type prediction. In practice, the inherent complexity of type definitions makes type recovery on binary code a highly challenging task. {In this work, we introduce several tradeoffs and restrict our focus to a small set of target types. This decision is motivated by the highly imbalanced distribution of type information in binary code: a small subset of types already accounts for the vast majority of type instances. Moreover, concentrating on this subset allows us to collect sufficient high-quality samples for each type during training, enabling our model to produce accurate and reliable prediction results.}

\subsection{Binary Code Semantics Representation \label{sec:binarycodesemanticsrepresentation}}

During the compilation process, the transformation from source code to binary code results in the loss of structural information, such as function boundaries and branch transitions, as well as semantic information including function names and variable names. This inherent information loss presents unique challenges in binary code analysis, leading to the emergence of specific tasks such as Binary Code Similarity Detection (BCSD) \cite{yang2023towards, yang2023Asteria-Pro}, variable type prediction \cite{zhu2024tygr, chen2021augmenting}, and function name prediction \cite{han2021issta, song2024bin2summary}. These tasks invariably require an accurate semantic representation of binary code. Logically, there are typically two main modeling paradigms for representing the semantics of binary code: sequence modeling and graph modeling.

\textbf{Sequence Modeling.} Binary code can be represented as assembly instructions or pseudo code, which inherently take a sequential form similar to natural language, making them suitable for sequence modeling. Tokenization presents the primary technical challenge due to binary instructions' fine-grained structural components, which include opcodes, operands, and other auxiliary tokens. The current leading method, PalmTree \cite{li2021palmtree}, employs a rule-based approach, treating each instruction as an independent statement unit for segmentation. Considering that sequence representation struggles to fully preserve structural information such as program control flow, PalmTree enhances its semantic understanding of binary code by integrating multiple pre-training tasks, including Masked Language Modeling (MLM) and Context Window Prediction (CWP). {Since assembly instructions lack comprehensible semantic information, CLAP \cite{wang2024clap} was the first to propose using natural language supervision to learn semantic representations of binary code, connecting natural language with binary code.} The resulting embedding vectors serve as semantic representations of binary code, providing foundational support for downstream tasks.

\begin{wrapfigure}{!t}{0.61\textwidth} 
    \vspace{-4mm}
    \begin{center}
        \subfigure[ByteTR's Graph]{
            \includegraphics[width=0.30\textwidth]{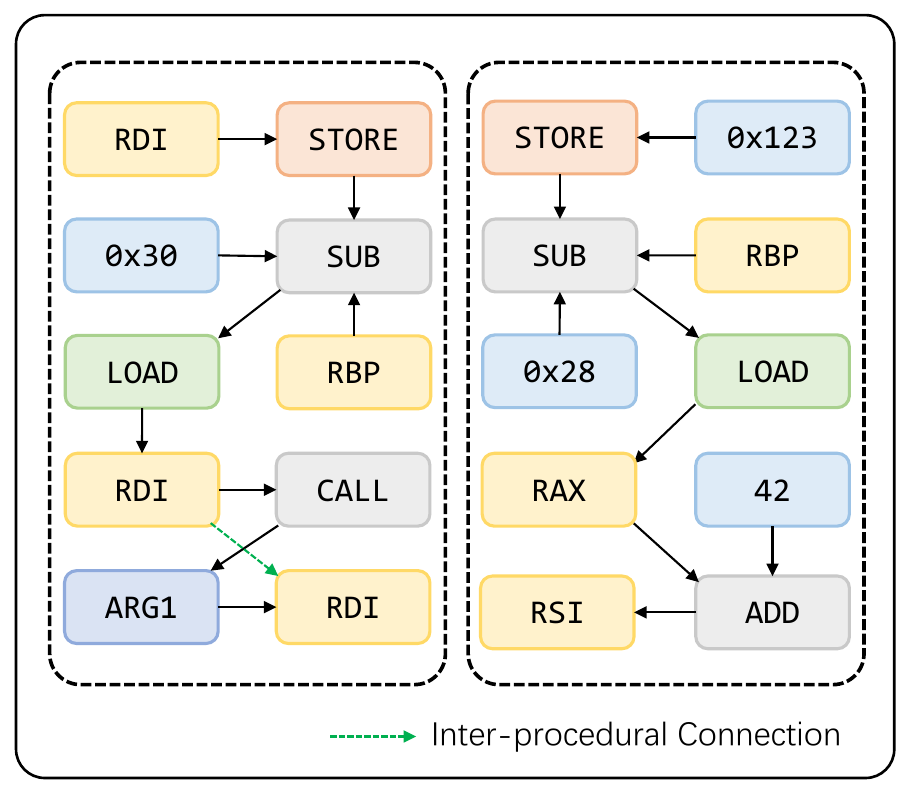}
            \label{fig:ByteTR_compare}
        }
        \hspace{0.00\textwidth}
        \subfigure[TYGR's Graph]{
            \includegraphics[width=0.26\textwidth]{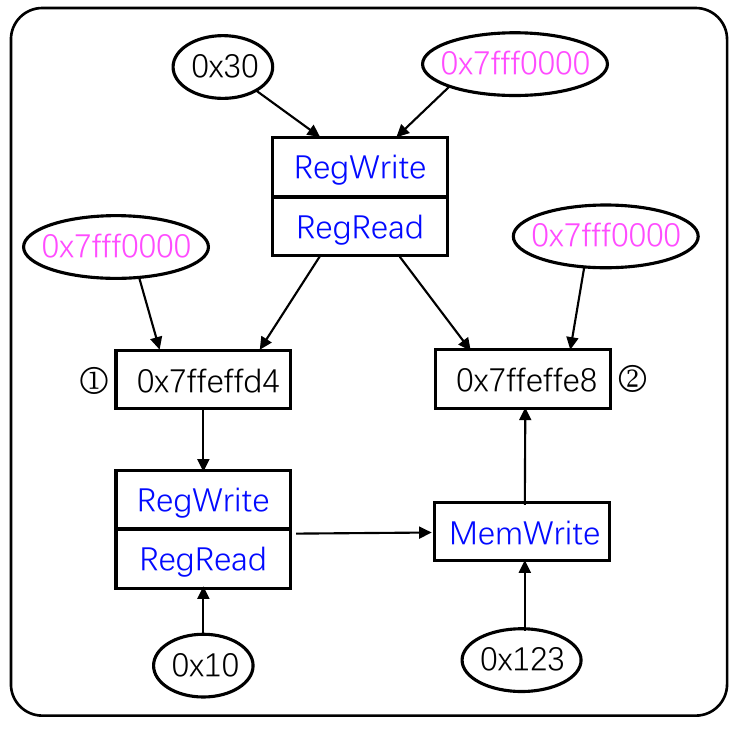}
            \label{fig:Tygr_compare}
        }
    \end{center}
    \vspace{-1em}
    \caption{{Differences in Graph Modeling between ByteTR and TYGR}}
    \label{fig:compare_in_graph_modeling}
    \vspace{-1em}
\end{wrapfigure}

\textbf{Graph Modeling.} Binary code also exists in graphical forms, such as control flow graphs (CFG) of assembly code and Abstract Syntax Tree (AST) of pseudo code, where the topological features provide rich structural information. IRBinDiff \cite{shang2024binary} embeds each token in a basic block and then aggregates them into the embedding CFG node, using the jump relationships as edges to generate the embedding of the entire graph. For the AST, DIRE \cite{lacomis2019dire} decouples each AST node into three components: syntactic type, data type, and node name, using their concatenation as the initial node embedding, and learns the representation of the entire graph during the training process.

While both our method and TYGR employ graph modeling to recover variable types, they differ fundamentally in their construction and granularity of analysis. As illustrated in Figure \ref{fig:compare_in_graph_modeling}, TYGR focuses on memory read and write features, building a function-level graph through symbolic execution to capture these operations. Its nodes partially  consist of concrete values computed during execution. In contrast, ByteTR utilizes static program analysis to decouple variables from functions, creating a variable-level propagation graph. This graph depicts the complete lifecycle of a variable from its definition to its use, with its nodes being more fine-grained instruction semantic units. Furthermore, the graph constructed by TYGR only describes intra-procedural behavior, whereas our graph can effectively build inter-procedural connections between variables.

\section{Empirical Analysis \label{sec:empiricalanalysis}}

Given our analysis of the variable type recovery dilemma in Section \ref{sec:introduction}, in this section, we conduct an empirical study to explore the properties of variable types in binary code. These empirical observations can help us to correctly formalize the task of variable type recovery in binary code so that we can have a better solution. We first present our research objectives and analytical approach. Subsequently, we describe the experimental setup and dataset preparation to ensure reproducibility. Finally, we analyze the experimental results and summarize our findings.

\subsection{Study Design \label{sec:empiricalanalysis.studydesign}}

To conduct an in-depth analysis of variable type characteristics in binary code, and unravel patterns for variable type prediction using deep learning techniques, we analyze this problem through three primary perspectives: Type Property, Variable Property, and Compilation Effects.

\subsubsection{\textbf{Type Property.}} Types in binary files are stored in the separate .debug section via the DWARF debugging format. Although physically isolated from the binary code, they are mapped through the symbol table. This design establishes a dynamic association between machine-level variables and the high-level language type system. Type Property characterizes the inherent attributes of variable types in binary code. While we have analyzed C language's type system from a language specification perspective in Section \ref{sec:background}, empirically investigating variable types in binary code enables a deeper understanding and better modeling of this problem in real-world scenarios. 

We investigate type properties through the type diversity, which reflects the richness and variety of variable types in binary code, alongside the abstraction of their corresponding variables. Specifically, we analyze type diversity using Type Frequency, Zipf’s Law, and Heaps’ Law, {adopting a methodology inspired by natural language processing. By analogy to linguistic tokens and documents, we model each type as a token and the aggregated type sequences within a binary as a document, thereby enabling quantitative analysis of their statistical distributions.}

\textbf{Type Frequency} represents how often different type tokens appear in the {debug information} of binary {programs} corpus. This attribute reflects which types of variables programmers prefer to use, and also indicates that these are the core types we should focus on for type recovery.

\textbf{Zipf's and Heaps' Laws} were initially observed in natural language text corpora \cite{piantadosi2014zipf, powers1998applications, sano2012zipf} as empirical laws, and were later proven to exist in programming languages as well \cite{zhang2008exploring, zhang2009discovering}. Zipf's law states that in human-generated natural text corpora when word frequency lists are arranged in descending order, the frequency value of the $n$-th entry is approximately inversely proportional to $n$. Subsequently, Zipf's law was extended to a general form as
\begin{equation}
\text{Frequency} \propto \frac{1}{(\text{Rank} + b)^a}
\label{eq:zipf}
\end{equation}
where $a\approx1$ and $b$ is a fitted parameter. In programming languages, Zipf's law explains that human-generated natural texts, such as function names, variable names, and method names, are mostly composed of a small number of frequently used tokens.

Heaps' law is another empirical law based on frequency, which describes the relationship between the {number of unique words} $V$ and the {document length} $n$ in a corpus as
\begin{equation}
V(n) = Kn^\beta
\label{eq:heaps}
\end{equation}
where $K$ and $\beta$ are fitted parameters. Empirically, Heaps' law is preserved even when documents are randomly shuffled, suggesting that it is independent of word order and depends solely on word frequency. Heaps' law implies that as more text is collected, the rate of encountering different vocabulary items decreases. In programming languages, Heaps' law typically applies to program entities such as function names and variable names.

Investigating the statistical patterns of variable types in binary code through the perspective of Zipf's law and Heaps' law provides insights into type distributions within datasets, thereby enabling more accurate approaches to modeling the type prediction problem.

\subsubsection{\textbf{Variable Property.}} 

The program behavior of variables reflects their intrinsic properties. Although different variables assume different roles in a program, this study focuses on their common properties in order to obtain valuable observations. Specifically, we provide an analysis of the properties of variables in terms of two perspectives: Variable Propagation and Locality of Reference of variable access.

In program execution, the propagation of local variables occurs when these variables are transmitted as function arguments to other functions, representing a fundamental phenomenon in program behavior. Previous research \cite{zhu2024tygr, chen2021augmenting, lacomis2019dire} has primarily focused on program slicing within individual functions for variable recovery, resulting in suboptimal performance. Intuitively, variable behavior manifests along its propagation path, and the accumulation of features contributes positively to variable type determination. We analyze the characteristics of variable propagation from two perspectives: Number of Functions reachable by a single variable, and Number of Variables passed as arguments within a function.

\textbf{Number of Functions} indicates the propagation scope of a variable, where a broader function scope introduces more characteristics of variable types.

\textbf{Number of Variables} indicates how many variables within a function propagate to other functions through variable propagation. This also reveals the patterns of variable propagation from an alternative perspective.

Locality of Reference states that processors tend to access the same set of memory locations repeatedly over a short period of time, a phenomenon also known as the principle of locality \cite{denning2005locality}. The locality principle consists of two basic types: temporal locality and spatial locality. Temporal locality describes the fact that the same memory location is more likely to be accessed multiple times in a relatively short period of time, while spatial locality indicates that memory regions close to the current access location are more likely to be accessed in {a near future}. In this study, since variables of primitive types usually occupy only one memory cell, we choose to explore the locality principle for variables of \texttt{struct} types, since structures usually contain multiple members and can better represent the locality principle in practice.

\textbf{Locality of Reference} characterizes the access patterns of the members of a structure variable in the temporal and spatial dimensions.

\subsubsection{\textbf{Compilation Effects.}}

Compilation transforms {human-readable} source code into {machine-readable} binary code, and compilers generate binaries with different execution behaviors through various optimization options to accommodate different platforms. In the open-source compiler, GCC \cite{gcc}, O0 is the default optimization option (i.e., do not optimize), where the compiler's objective is to reduce compilation costs and ensure debugging produces the expected results. O1 is the recommended optimization level for large machines, serving as a reasonable balance between compilation time and memory usage. O2 and O3 further optimize code performance but increase compilation time without considering the space-speed trade-off. Compiler optimizations generate code with enhanced performance, which implies reducing instruction count and utilizing more complex instructions. This inevitably affects variable behaviors and semantic representations \cite{ren2021unleashing, jiang2024bincola}, making variable recovery more challenging. 

\textbf{Pattern of Storage} of variables typically follows a stack-based Load-Compute-Store mode without optimizations, where variables are stored in function stack frames. During execution, variables are loaded from memory into registers, computations are performed, and results are subsequently stored back into memory. Compilation optimizations reduce frequent memory accesses and use registers to store the value of a variable, thus affecting the way the variable is stored, changing it to be register-based. When a function accesses a global variable, since the function does not own the variable, the access pattern is usually address-based. Different storage patterns manifest distinct program behaviors, consequently impacting variable recovery.

\subsection{Experimental Setup \label{sec:empiricalanalysis.experimentalsetup}}

Variable type data in binary files are typically stored as unidirectional linked \texttt{DW\_AT\_type} entries, where each variable entry contains the head node of its type linked list. We parse these type linked lists to reconstruct complete variable types, which enables us to analyze type frequency distributions and further validate their conformance to Zipf's and Heaps' Laws.

We employ the IDA {Pro} decompiler \cite{IDA_PRO} to analyze the unidirectional relationships in binary files, specifically tracking variables passed as arguments between functions. These unidirectional relationships are utilized to construct a directed graph that models inter-procedural variable propagation patterns. Through depth-first search traversal, we compute the reachability of individual variables to determine both the Number of Functions and Number of Variables within the propagation path. Since the binary dataset does not provide a mapping at the source code level, we also use the IDA Pro decompiler tool to generate pseudo code and statistically analyze the referencing patterns of the structure members in it to explore the Locality of Reference. 

The location expressions of variables are stored in the DWARF information of binary files. We analyze the location expressions to determine the Pattern of Storage of variables to study Compilation Effects. 

All experiments were conducted on a Linux server equipped with an 112 logical cores' Intel Xeon Gold 6330 CPU and 8 NVIDIA RTX A6000 GPUs.
 
\subsection{Dataset Preparation \label{sec:empiricalanalysis.datasetpreparation}}

Deep Learning-based models heavily rely on high-quality datasets for training and evaluation. Previous studies \cite{pei2021stateformer, lacomis2019dire, chen2021augmenting} on type recovery have revealed that their released datasets exhibit a high function duplication rate, primarily attributed to the compilation of different versions of the same software (e.g., binutils 2.43, binutils 2.44), which have minimal source code variations. Notably, StateFormer and DIRTY demonstrated remarkably high function duplication rates of 89.9\% and 65.5\%, respectively \cite{pal2024len}. Recent work TYGR \cite{zhu2024tygr} systematically analyzed the limitations of previous datasets and introduced TYDA, a binary dataset comprising 163,643 binary programs sourced from C packages in Gentoo and Debian repositories. TYDA has been peer-reviewed for its superior volume and quality compared to previous datasets.

By leveraging pyelftools to parse DWARF information from TYDA binary programs, we obtained 21 million functions and 105 million variables from x86-64 architecture binaries compiled with O0 optimization level alone. 

\subsection{Results and Analysis \label{sec:empiricalanalysis.resultsandanalysis}}

We conduct an empirical experiment and comprehensive analysis of each point mentioned in Section \ref{sec:empiricalanalysis.studydesign}, aiming to explore the inherent properties of type recovery tasks. {Before delving into the specific analysis, we first present a clear mind map of our entire study. Finding 1 and 2 investigate the natural distribution of types in binary programs, analyzing them from the perspectives of frequency and of Zipf's and Heaps' Laws, respectively. Finding 3 focuses on a special complex type, the structure, to explore the access patterns of its members. Shifting the focus from types to variables, Finding 4 examines their propagation features across procedural boundaries. Finally, Finding 5 analyzes the impact of compiler optimizations on the storage patterns of variables.}

\begin{table}
  \caption{The top 50 variable types sorted by frequency in the TYDA dataset.}
  \vspace{-1ex}
  \label{tab:freq}
  \setlength{\tabcolsep}{1.5mm}
  \scalebox{0.83}{
  \begin{tabular}{ccl}
    \toprule
    Rank & Data Type \\
    \midrule
     1-10 & int, \colorbox{gray!30}{const char}, char *, double, void *, PyObject *, \colorbox[rgb]{0.792,0.914,0.682}{size\_t}, unsigned int, long int, int * \\
    11-20 & uint32\_t, double *, uae\_u32, object, gpointer, float, \colorbox{gray!30}{const gchar *}, gint, \colorbox{gray!30}{const int}, char ** \\
    21-30 & pair\_type, uint8\_t *, \colorbox[rgb]{1.000,0.847,0.576}{\_Bool}, uint8\_t, \colorbox{gray!30}{const void *}, \colorbox[rgb]{1.000,0.847,0.576}{gboolean}, obj\_t, FILE *, uint64\_t, Py\_ssize\_t\\
    31-40 & guint, gchar *, object *, GtkWidget *, uint, long int *, const uint8\_t *, char,\colorbox[rgb]{1.000,0.847,0.576}{bool}, \colorbox[rgb]{0.792,0.914,0.682}{long unsigned int}\\
    41-50 & float *, closureN\_type, int32\_t, \colorbox{gray!30}{const uchar *}, intnat, GError **, int64\_t, mlsize\_t, unsigned char, uint32\\
  \bottomrule
  \end{tabular} 
}
\vspace{-1.5ex}
\end{table}

\subsubsection{\textbf{Type Frequency}} Table \ref{tab:freq} presents the top 50 data types in the dataset, ranked by their frequency of occurrence. Based on these results, we have several noteworthy observations.

\textbf{Primitive Types}: Empirically, the most frequently used type is the \texttt{\textbf{int}} type, which intuitively aligns with it being the first data type learned in programming education. Furthermore, among the top 50 data types, 26 are primitive types (e.g., \texttt{char}, \texttt{double}), indicating that primitive types are the core types in binary datasets. This suggests that in C programs, developers tend to favor primitive types for expressing logic. Therefore, we should prioritize primitive types in binary type recovery tasks due to their substantial proportion in the dataset.

\textbf{Type Qualifier}: Keywords such as \texttt{const}, \texttt{volatile}, \texttt{restrict}, and \texttt{static} in types are used to constrain type characteristics and instruct the compiler to perform specific checks. The \texttt{const} qualifier declares a read-only variable, more specifically, the compiler verifies that the variable cannot be directly assigned after initialization. However, a variable that naturally remains unassigned after initialization cannot be distinguished at the binary code level from a const-qualified variable. Currently, in the reverse engineering process of analyzing variables and inferring types, we do not focus on these type qualifiers. In Table \ref{tab:freq}, there are 5 types with the \texttt{const} type qualifier (highlighted in gray), and their ground truth types are the types with the qualifiers removed.

\textbf{Type Aliasing}: In the Table \ref{tab:freq}, \texttt{size\_t} is an alias type defined through \texttt{typedef} for \texttt{long unsigned int} (highlighted in green background). From the logical perspective of programming, \texttt{size\_t} is used to store an object's size variable, and its unsigned integer nature can be inferred from the word "size". However, in most scenarios without prior knowledge, it is impossible to deduce the original type of an alias type, such as \texttt{pid\_t}, \texttt{pair\_type}. Type aliasing also leads to type duplication, as seen with \texttt{\_Bool}, \texttt{gboolean}, \texttt{bool} (highlighted in yellow background), which are fundamentally defined based on \texttt{\_Bool} despite their different names. Such duplication results in an expansion of the type vocabulary.

\begin{tcolorbox}[colback=gray!5,
                  colframe=black,
                  arc=0.8mm, auto outer arc,
                  boxrule=1pt,
                  boxsep=-2pt
                 ]
\textbf{Finding 1:} Primitive types are the core of the type system, appearing with significantly higher frequency and proportion in real-world collected datasets. Furthermore, the type vocabulary exhibits extensive duplication, where Type Qualifier and Type Aliasing create types that are identical in nature but differ in names.

\end{tcolorbox}

\begin{figure}[!t] 
    \centering
    \subfigure[Zipf's Law ($a=1.1, b=0, p<0.001$) ]{
        \includegraphics[width=0.45\textwidth]{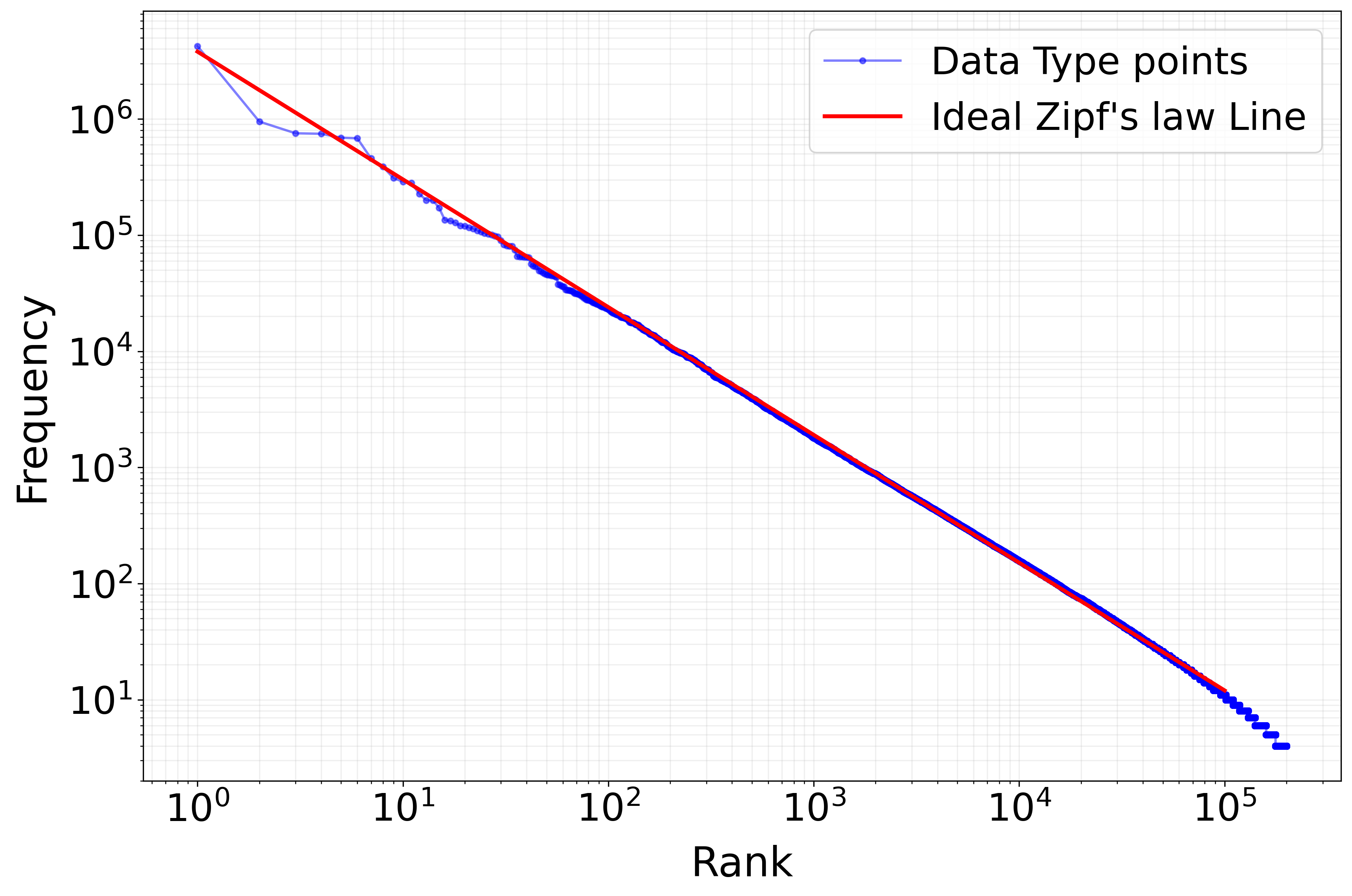}
        \label{fig:zipfs.law}
    }
    \subfigure[Heaps' Law ($K=1.41, \beta=0.7, p<0.005$)]{
        \includegraphics[width=0.45\textwidth]{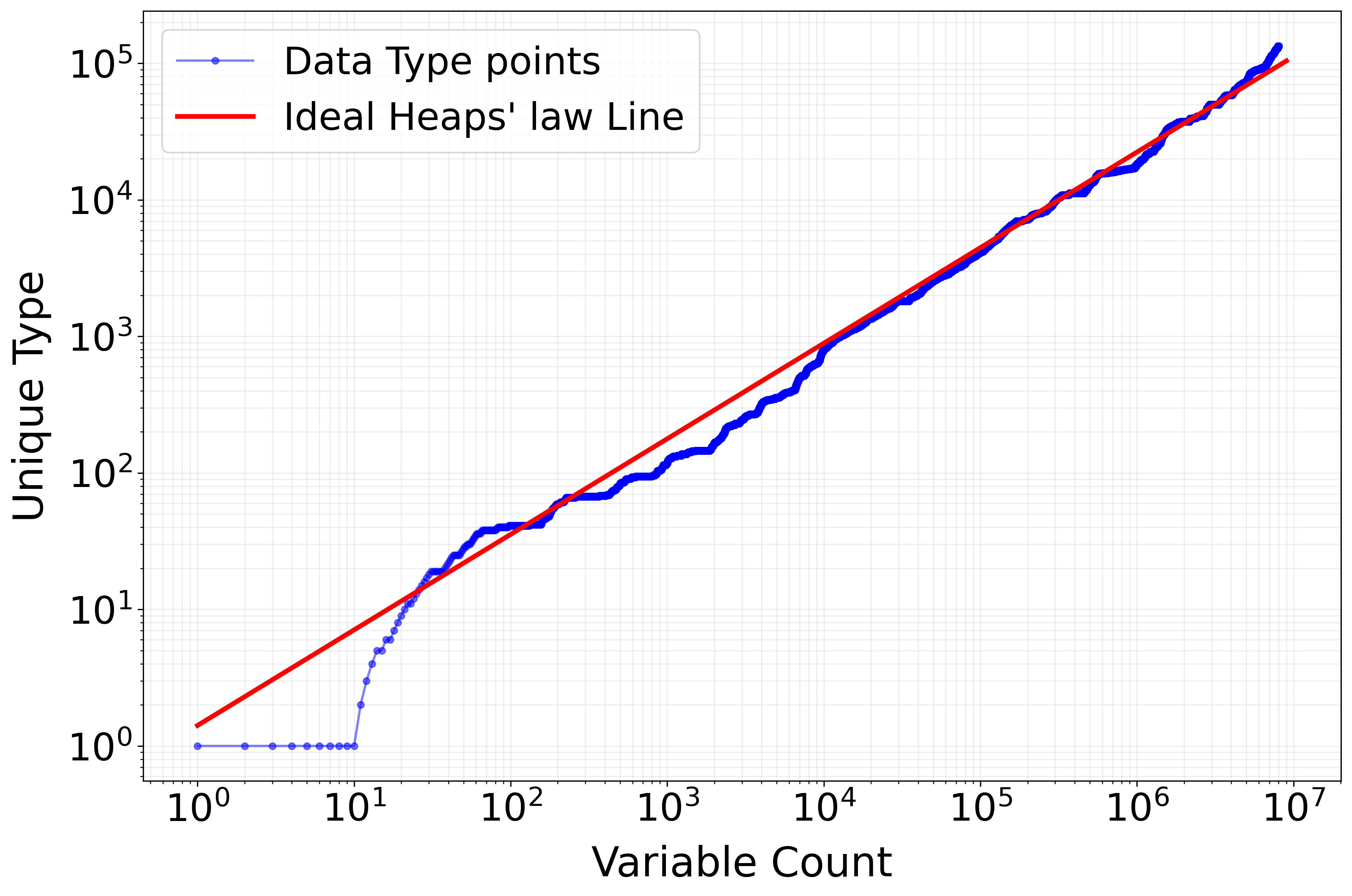}
        \label{fig:heaps.law}
    }
    \vspace{-2ex}
    \caption{Variable type data fitting Zipf's and Heaps' laws where $p$ denotes statistical significance.}
    \label{fig:zipfs.and.heaps.law}
    \vspace{-1ex}
\end{figure}

\subsubsection{\textbf{Zipf’s and Heaps’ Laws}} Figure \ref{fig:zipfs.and.heaps.law} provides a comprehensive understanding of the natural distribution of types. It demonstrates that the type data fits Zipf's law with a statistical significance $p$ < 0.001 and Heaps' law with $p$ < 0.005. Based on these empirical laws, we can draw several significant insights.

\textbf{Unbalanced Distribution:} Zipf's law directly demonstrates that the word frequency of natural types is inversely proportional to their rank, indicating a severely unbalanced distribution in type data. In practice, high-frequency types, which constitute approximately 20\% of the total type categories, account for 80\% of the actual use cases. 

\textbf{Increasing Vocabulary:} Heaps' Law demonstrates that as the number of type instances increases, the count of unique types also grows, but at a decreasing rate ($\beta < 1$). In natural language, this law has informed the design of tokenizer for large language models, which preserves high-frequency words while decomposing low-frequency ones into subwords. However, due to the atomic nature of types, the strategy of generating an appropriate vocabulary by splitting into subwords is not feasible. The continuously increasing vocabulary, as predicted by Heaps' Law, remains an unresolved challenge in this domain.

\begin{tcolorbox}[colback=gray!5,
                  colframe=black,
                  arc=0.8mm, auto outer arc,
                  boxrule=1pt,
                  boxsep=0pt
                 ]
\textbf{Finding 2:} Type's frequency distribution precisely follows both Zipf's Law and Heaps' Law. According to Zipf's Law, the frequency of types exhibits an inverse relationship with their frequency ranking, indicating a significant unbalanced distribution in the type data. According to Heaps’ Law, as token instances increase, unique types also grow, but at a slower rate, reflecting a gradual expansion of the type vocabulary.
\end{tcolorbox}

\subsubsection{\textbf{Locality of Reference}} Table \ref{tab:member.reference} summarizes member variable references extracted from binary pseudo code. Unlike high-level languages, binary structure access relies on memory addressing: a base address, a specific Offset, and a read size (Bytes). Hits reflects total access instances, while Unique indicates the count of distinct structures referenced.

\begin{table}[H]
    \centering
    \caption{Member Reference of structures in pseudo code of TYDA binaries.}
    \vspace{-3ex}
    \label{tab:member.reference}
    \setlength{\tabcolsep}{1.9mm}
    \scalebox{0.83}{
    \begin{threeparttable}
    \begin{tabular}{@{}wc{1.3cm}|cccc||wc{0.9cm}|cccwc{1.7cm}@{}}
        \toprule
        \textbf{Rank} & \textbf{Offset} & \textbf{Bytes} & \textbf{Hits} & \textbf{Unique\tnote{1}} & \textbf{Rank} & \textbf{Offset} & \textbf{Bytes} & \textbf{Hits} & \textbf{Unique} \\ 
        \midrule
        1  &  0 &   8 &   15,879,630 &   1,056,220  & 11 &  0 &   1 &     964,439 &      45,840  \\
        2  &  8 &   8 &    6,970,183 &     379,603  & 12 & 40 &   8 &     937,370 &      70,608   \\
        3  &  0 &  24 &    5,116,204 &     186,791  & 13 &  0 &  32 &     925,023 &      42,885   \\
        4  &  0 &   4 &    3,395,816 &     142,196  & 14 &  0 &   2 &     902,189 &      14,749   \\
        5  & 16 &   8 &    3,272,634 &     252,790  & 15 & 48 &   8 &     855,166 &      62,947  \\
        6  & 24 &   8 &    2,097,416 &     156,895  & 16 &  0 &  48 &     781,449 &      56,973  \\
        7  &  0 &  16 &    1,869,511 &     102,909  & 17 & 16 &   4 &     774,460 &      56,835 \\ 
        8  &  8 &   4 &    1,543,342 &      99,469  & 18 & 56 &   8 &     623,672 &      46,574  \\ 
        9  & 32 &   8 &    1,314,585 &     104,077  & 19 &  0 &  80 &     584,338 &      15,253  \\ 
        10 &  4 &   4 &    1,219,549 &      72,386  & 20 & 12 &   4 &     578,346 &      47,907  \\ 
        \bottomrule
    \end{tabular}
    \begin{tablenotes}
    \item[1] Number of unique structure types.
    \end{tablenotes}
    \end{threeparttable} }
    \vspace{-1ex}
\end{table}

\textbf{Member Reference Locality}. Focusing on the data in the Offset and Bytes columns, as shown in Table \ref{tab:member.reference}, it can be found that the memory cell with an offset of 0 has the highest access frequency, and the members of its neighboring memory cells are also ranked high in terms of access frequency. This phenomenon suggests that the accesses of structure members are not uniformly distributed, but show obvious localization characteristics. Specifically, only some key members (e.g., members of rank1 and rank2) are frequently accessed during the lifecycle of structure variables, while other members are relatively idle. Actually, in a type recovery task, it is not possible to recover all the members of a structural variable when not every member leaves a trace.

\textbf{Memory Overlap.} As shown in Table \ref{tab:member.reference}, millions of structure variables store members of the same type at the same offset location, and notably, this memory overlap is prevalent in the top 20 structures we observed. According to the Locality of Reference principle, members in these shared locations tend to have higher access frequencies. Therefore, from the perspective of a single structural variable, when a structural variable whose scope is limited to a certain function only accesses the 8-byte member located at offset 0 during its lifetime, the variable may correspond to millions of potential structural types due to the memory overlap phenomenon, which makes it nearly impossible to accurately recover its original type definition. This phenomenon also reveals that the essence of a structure lies in the way its memory is laid out and its members are organized, rather than in the way it is named at the time of its definition.

\begin{tcolorbox}[colback=gray!5,
                  colframe=black,
                  arc=0.8mm, auto outer arc,
                  boxrule=1pt,
                  boxsep=-2pt
                 ]
\textbf{Finding 3:} The access pattern of structure members shows a significant Locality of Reference: only a few key members are frequently accessed during the variable's life cycle, while others are relatively unused. These high-frequency accessed members usually represent common properties, resulting in the tendency of different structures to use the same data type at the same offset location, forming Memory Overlap. 
\end{tcolorbox}

\begin{figure}[!t]
    \centering
    \subfigure[Number of Functions]{
        \includegraphics[width=0.46\textwidth]{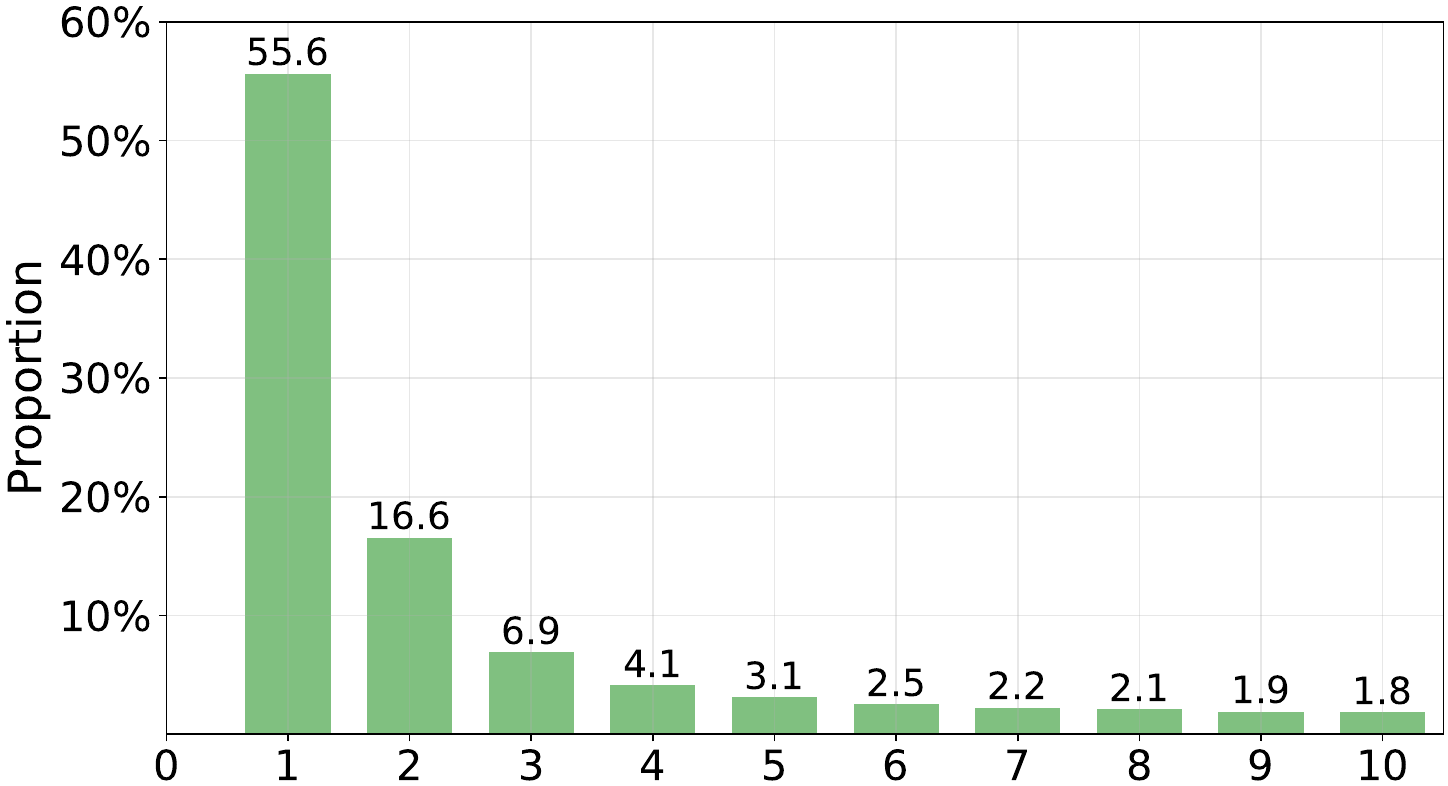}
        \label{fig:number.of.function}
    }
    \subfigure[Number of Variables]{
        \includegraphics[width=0.43\textwidth]{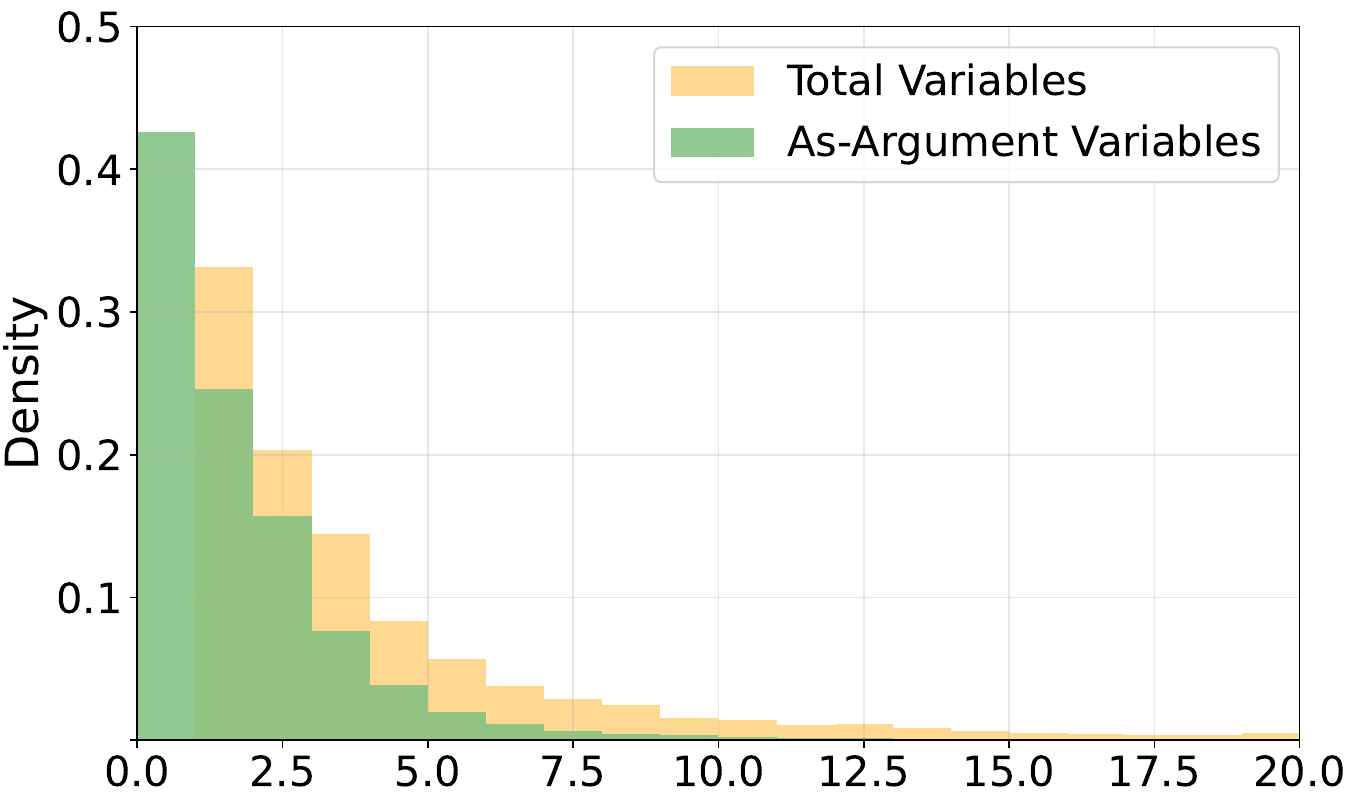}
        \label{fig:number.of.variable}
    }
    \vspace{-2ex}
    \caption{Statistical analysis of variable propagation. (a) Proportion of the number of functions crossed by a single variable as it propagates along the data flow. 
    (b) Density distribution of the total number of variables {per function and those propagated as arguments to other functions}.
    }
    \vspace{-2ex}
    \label{fig:number.of.function.and.variable}
\end{figure}

\subsubsection{\textbf{Number of Functions and Variables}} The analysis of variable propagation, as visualized in Figure \ref{fig:number.of.function.and.variable}, reveals insights into the function call. The Number of Functions and Number of Variables both show a gradually decreasing trend, conforming to a right-skewed distribution. Figure \ref{fig:number.of.function} demonstrates the proportion of variables from the perspective of functions, showing that 55.6\% of the variables have a scope limited to a single function, while implying that the remaining 44.4\% of the variables propagate across functions. Nearly half of the variables propagate across multiple functions demonstrates the need for inter-procedural analysis when characterizing the semantics of variables. The behavior of variables between functions can effectively enhance the characterization of variables. As the number of functions increases, the proportion of variables decreases while remaining relatively stable. We can analyze the existing stable variables, which in practice are often used as object pointers and passed as parameters to each function to provide semantic context.

We demonstrate the function's distribution from the perspective of variables to reveal the patterns of variable propagation in Figure \ref{fig:number.of.variable}. First, it examines the distribution of functions with varying numbers of total variables, as represented by the yellow bins in the figure. Most functions contain ten or fewer total variables, while the density of functions with more than twenty variables is very low. This phenomenon suggests that C language functions in real-world projects tend to be smaller in size, utilizing a limited number of variables to implement clear and concise logic, which aligns with the fundamental design principle of decoupling in software engineering.

In addition, regarding the function's distribution of the number of variables that are propagated across functions, the green bins indicate that approximately 40\% of the functions do not contain any variables propagated across functions, while the remaining functions contain at least one such variable. Overall, the green distribution is a leftward shift of the yellow distribution, suggesting that a significant proportion of variables within a single function propagate to other functions. 

\begin{figure}[!t]
    \centering
    \includegraphics[width=0.9\textwidth]{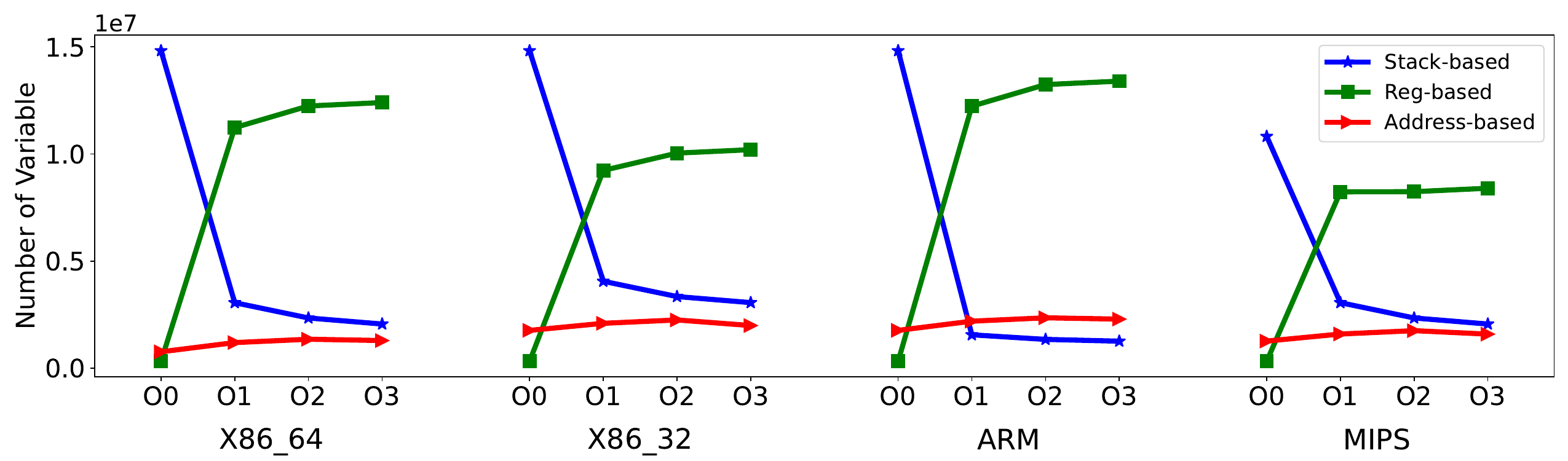}
    \vspace{-1.7ex}
    \caption{Variable storage patterns across different architectures and optimization options}
    \vspace{-3ex}
    \label{fig:pattern.of.storage}
\end{figure}

\begin{tcolorbox}[colback=gray!5,
                  colframe=black,
                  arc=0.8mm, auto outer arc,
                  boxrule=1pt,
                  boxsep=-2pt
                 ]
\textbf{Finding 4:} Variable propagation across function boundaries emerges as a prevalent pattern in real-world datasets. From the perspective of variables, nearly half of them are propagated as arguments via function calls. From the perspective of functions, close to 60\% of them contain variables that propagate across functions, suggesting that a significant proportion of variables propagate to other functions.
\end{tcolorbox}

\subsubsection{\textbf{Pattern of Storage}} We investigate the impact of compiler optimization on variable storage patterns. As shown in Figure \ref{fig:pattern.of.storage}, we conduct a systematic analysis of variable storage patterns across different architectures under various optimization levels. The results demonstrate that as the optimization increases, the count of stack-based variables exhibits a significant downward trend, while register-based variables increase correspondingly. Notably, the address-based variable remains relatively stable, showing minimal sensitivity to optimization levels. 

When the optimization is not turned on (O0), variables are primarily stored in stack-based pattern, while the O1 optimization reverses this by optimizing variables to a register-based pattern. Higher optimization options O2, O3 exacerbate this change. These compilation effects are presented on variables of different architectures, but with subtle differences. The compilation optimization stores variables into registers to reduce memory accesses, but due to limitations of ABI \cite{amd64-abi}, these registers must be callee-saved to ensure guaranteed correct data transfer. The ABI of the different architectures have a different number of callee-saved registers, making the situation slightly different, with more on ARM architecture and fewer on x86-32 architecture.

Different storage patterns for variables lead to different program behaviors, with stack-based storage variables exhibiting more memory accesses, while register-based ones do not. Thus, approaches based on dynamic program analysis \cite{zhu2024tygr} that capture accesses to variables to model program semantics have a harder time coping with register-based variables. 

\begin{tcolorbox}[colback=gray!5,
                  colframe=black,
                  arc=0.8mm, auto outer arc,
                  boxrule=1pt,
                  boxsep=-2pt
                 ]
\textbf{Finding 5:} Compiler optimization transforms the storage pattern of variables from stack-based to register-based, with this transformation being most significant at the O1 optimization level. These compilation effects are generally consistent across different architectures, with slight variations due to ABI influences.
\end{tcolorbox}

\section{Methodology \label{sec:methodology}}

\begin{figure}[!t]
    \centering
    \includegraphics[width=0.98\textwidth]{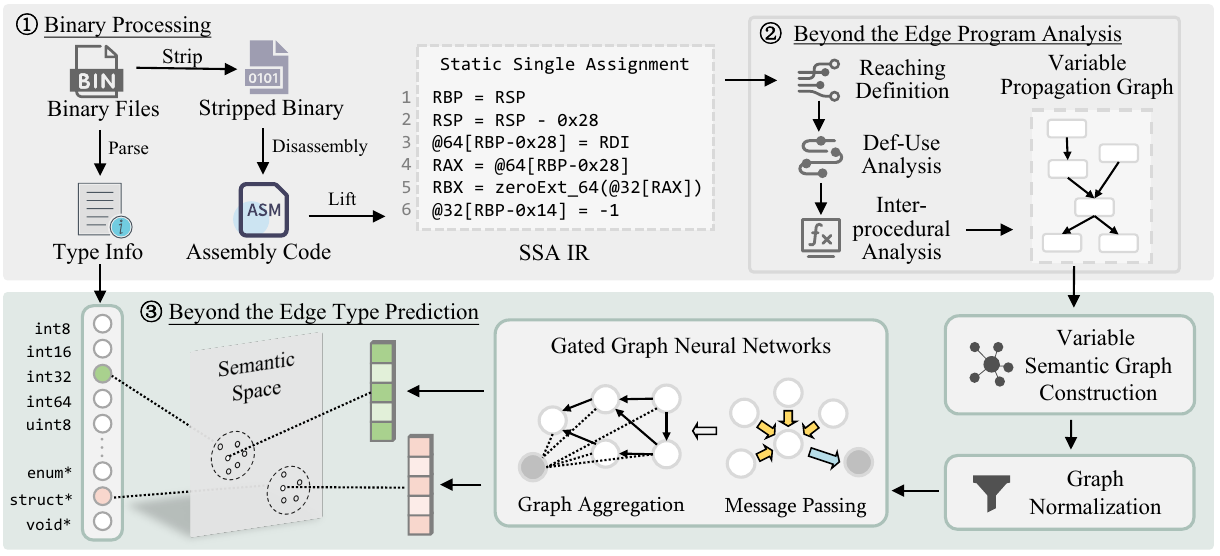}
    \vspace{-1.5ex}
    \caption{Overview of ByteTR.}
    \vspace{-2ex}
    \label{fig:overview.ByteTR}
\end{figure}

Based on extensive exploratory experiments in the previous section, we reveal empirical insights into type property, variable property, and the impact of compilation optimizations on variable behaviors. However, some existing type recovery methods exhibit contradictions with our empirical analysis results. For instance, nearly all prior works, such as DIRTY, TYGR, and StateFormer, do not support inter-procedural variable feature analysis, which directly contradicts our Finding 4. TYGR only supports predictions for stack-based variables but fails to handle register-based variables, opposing Finding 5. We argue that an ideal type recovery system should be fully grounded in empirical research when designing its methodology. Inspired by these insights, we develop an innovative technical approach ByteTR. We present the overview of ByteTR in Figure \ref{fig:overview.ByteTR}, which consists of two main components: (1) the novel BytePA (Beyond the Edge Program Analysis) algorithm, which constructs variable propagation graphs to capture fine-grained variable features through inter-procedure analysis, and (2) the ByteTP (Beyond the Edge Type Prediction) framework that employs GGNN to represent the semantics of variables and facilitate type recovery.

We start by formulating the variable type recovery problem in binary code, followed by a step-by-step introduction to each component of ByteTR.

\subsection{Problem Formulation \label{sec:methodology.problemformulation}}

We assume that the binary code is generated normally by the compiler and strictly follows the corresponding target ISA specification, ABI specification, and operating system specification. On this basis, we formalize the type recovery problem in binary code as a classification task whose central goal is to learn a mapping function $\Gamma(\cdot)$ that accurately predicts the variable types $T$ corresponding to variables $V$.
\begin{equation}
\Gamma(V) = T
\label{eq:Methodology.Formulation1}
\end{equation}

The type of a variable not only gives the variable a specific semantics, but also restricts its behavior, which is reflected in the semantics of the binary code fragments that operate on the variable. {Since variable propagation is itself a data flow, we choose to use the Graph Modeling and }employ a directed graph $G$ to represent the semantics of the {variable}. {Due to the natural directedness of data flow, which does not loop back on itself, we define the edges $E$ as pairs$(u_1, u_2)$  and use $u_1 \neq u_2$ to indicate that there are no self-loops.}
\begin{equation}
V = G(U, E), \quad E \subseteq \{(u_1, u_2) \mid u_1, u_2 \in U, u_1 \neq u_2 \}
\label{eq:Methodology.Formulation2}
\end{equation}

{Our empirical analysis investigates the representation of variables in binary code, with Finding 5 highlighting the severe impact of compiler optimizations on variable storage. Using traditional assembly instruction sequences to model the variables would require designing complex pre-training tasks to learn data dependencies between instructions and the irrelevance of variable registers and memory allocation. It would also require large-scale pre-training to understand these variations.  In contrast, the directed graph $G$ constructed from nodes and edges can naturally represent data dependencies and unify the modeling of variables in both registers and memory. This makes it inherently robust to semantic changes caused by different optimization levels. Furthermore, since variable propagation is essentially a data flow, using a graph structure to model variable semantics is a natural and effective approach.}

Finding 4 reveals the common phenomenon of variable propagation between functions, for which we propose the BytePA algorithm for inter-procedural program analysis, which aims to track the propagation behavior of variables across multiple functions. The algorithm reveals how variables behave in complex contexts by integrating the behavior of the same variable across multiple functions into a single graph. For external calls to library functions, {although the internal logic of the callee function is unavailable, the call-site argument setup phase forms an observable interface that can still reveal key features of the variables.}

\begin{figure}[!t]
    \centering
    \includegraphics[width=0.98\textwidth]{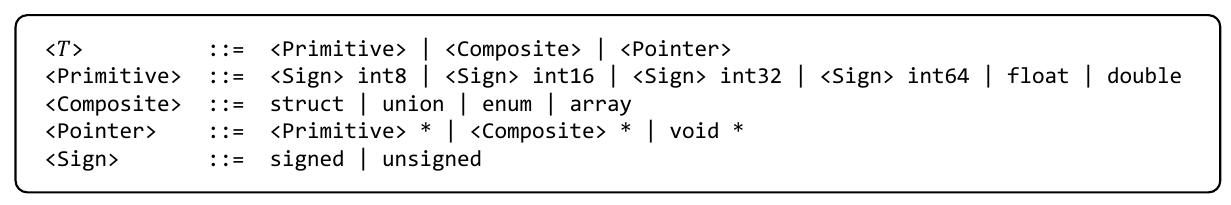}
    \vspace{-1ex}
    \caption{The full set of predictable types presented in the form of the BNF paradigm in our formulation.}
    \vspace{-1ex}
    \label{fig:supported.BNF}
\end{figure}

Finding 2 reveals the unbalanced distribution of types and that the type variety tends to increase and does not converge as the number of variable instances increases. Consequently, predicting all types is infeasible. Therefore, we choose to recover the types in $T$ that are the smallest unique type entities, as shown in Figure \ref{fig:supported.BNF}. In Finding 1, the primitive types have a high proportion of variable instances, and in practical reverse engineering, the base types play a crucial role in revealing the program semantics, so we decide to recover all primitive types in $T$. For alias types, since they have exactly the same characteristics as the original types, even in the compilation phase, the alias types are replaced with the original types to be analyzed for allocating registers and generating machine code. Therefore, in the type recovery process, we recover only the original types. 

Based on Finding 3, we observe that structure types exhibit Memory Overlap, which means that different structure types will use the same data types in the same memory locations. In addition, structure types exhibit Member Reference Locality, which means that in most cases, only {a subset of the} fields in the structure variable {are accessed with high frequency and diversity to reveal their characteristics}, while others will show {limited or no observable usage patterns}. As a result, the overall characterization of the structure is often not fully revealed, which makes it difficult to distinguish them from types with memory overlapping properties. Based on these observations, we decide to only determine whether a variable is a structure type without further predicting its specific identifier name, as shown in $T$. Accurately recovering the full structure of composite types is beyond the scope of the current research, and we will discuss this issue in detail in Section \ref{sec:discussion}.

We train the gated graph neural network $\Gamma(\cdot)$ to learn the features of the variables, which accepts inputs from the variable and predicts the corresponding data types.

\subsection{Beyond the Edge Program Analysis \label{sec:methodology.bytePA}}

In this section, we present the workflow of the BytePA algorithm, as shown in Figure \ref{fig:overview.BytePA}, and analyze step by step the process of how to decouple variables from the whole function, as well as tracking their behavior between functions, and ultimately generating Variable Propagation Graphs. 

\subsubsection{\textbf{Lifting to SSA IR}}
Static single assignment form (SSA) is a widely used intermediate representation during compilation, characterized by the fact that each variable is assigned only once. {It can be used for both high-level, architecture-independent IR (such as LLVM IR) and low-level IR that is tightly coupled with specific machine instructions.} Its concise form makes it ideal for optimizing code performance during compilation. {Although SSA IR was not originally designed to be architecture-agnostic, our approach uses the miasm \cite{miasm} framework to lift binary code into a unified SSA IR. This is crucial for enabling effective program analysis across different architectures.}

{We refer to the SSA IR lifted from binary code as $SSA_M$ IR, and contrast it with $SSA_S$ IR, which is from source code during compilation. Traditional program analysis algorithms are typically designed for $SSA_S$ IR. This is because $SSA_S$ IR variables directly correspond to source code variables and it contains explicit function call structures, which makes program analysis relatively straightforward. However, the subject of this work is $SSA_M$ IR. It naturally lacks the structural information from the source code, and its variables are registers or memory locations, which presents a significant challenge. To enable effective program analysis on $SSA_M$ IR, we have made substantial modifications and innovations to these traditional algorithms. Furthermore, to address the specific needs of inter-procedural variable connection, we have designed a new inter-procedural variable analysis algorithm. These approaches are fundamentally different from existing methods.}

\subsubsection{\textbf{Reaching Definition Analysis}}

Reaching Definition is a basic data-flow analysis methodology for determining where each variable $x$ is defined when the program execution flow reaches point $p$. Specifically, we say that definition $d$ of variable $x$ “reaches” point $p$ if there exists a path from definition $d$ to point $p$ and no other definition of $x$ on this path overrides $d$. Conversely, if there are other definitions of $x$ on the path, then definition $d$ will be “killed”. A typical application of Reaching Definition in compiler is to determine whether $x$ is a constant when the variable $x$ is used at point $p$, thereby optimizing code generation. {Compared to $SSA_S$ IR, registers or memory units in $SSA_M$ IR don't have a one-to-one correspondence with variables in the source code. Furthermore, $SSA_M$ IR also lacks information on the location of function parameters, which are crucial for inter-procedural analysis. To solve these problems, we will introduce an improved Reaching Definition method below.}

First, we decouple the real variables in the source code from the SSA$_M$ IR. In debug symbols, variables are described as location expressions, as shown in Figure \ref{fig:overview.BytePA}, the location of the variable \texttt{fd} is in memory cell \texttt{rbp-0x28}, while the variable \texttt{server\_fd} is stored in the \texttt{RBX} register.

We start the Reaching Definition analysis from the function start point $p$ of SSA$_M$ IR, and for each instruction analyzed, the definition position $d$ of the variable $x$ it defines is updated and the previous definition position is killed. When the point $p$ reaches 9, we give the result of the Reaching Definition as shown in Figure \ref{fig:overview.BytePA}, a table of the reached variables $V$ with the defined positions $D$. By looking up the expression of each variable in debug symbols in the analysis result, we are able to get the definition location $d = V[\hat{x}]$ of the variable $\hat{x}$. In this way, we can decouple the exact IR position of the variable definition from the entire function.

Based on the analysis of Finding 5, location expressions are susceptible to compilation optimization, and the higher the optimization level, the more variables tend to use register storage. Compared with the previous work \cite{zhu2024tygr}, which only supports the location of stack-memory variables, with the above analysis, our method can effectively deal with both stack-memory variables and register variables.

Next, we analyze function parameters. Unlike SSA$_S$ IR, in binary code, parameters are stored in specific registers according to the calling convention. Taking the x86-64 architecture as an example, the System V ABI \cite{amd64-abi} specifies that parameters are stored in the \texttt{RDI}, \texttt{RSI}, \texttt{RDX}, \texttt{RCX}, \texttt{R8}, and \texttt{R9} registers. To facilitate subsequent Def-Use analysis of parameter propagation, we introduce a dummy definition for each parameter at the entry point of SSA$_M$ IR to represent the assignment, as shown in Figure \ref{fig:overview.BytePA}.

Through Reaching Definition analysis, we successfully determine the definition locations of variables in SSA$_M$ IR. In addition, we add dummy definitions for each parameter to facilitate subsequent analysis.

\subsubsection{\textbf{Def-Use Analysis}}

\begin{figure}[!t]
    \centering
    \includegraphics[width=0.98\textwidth]{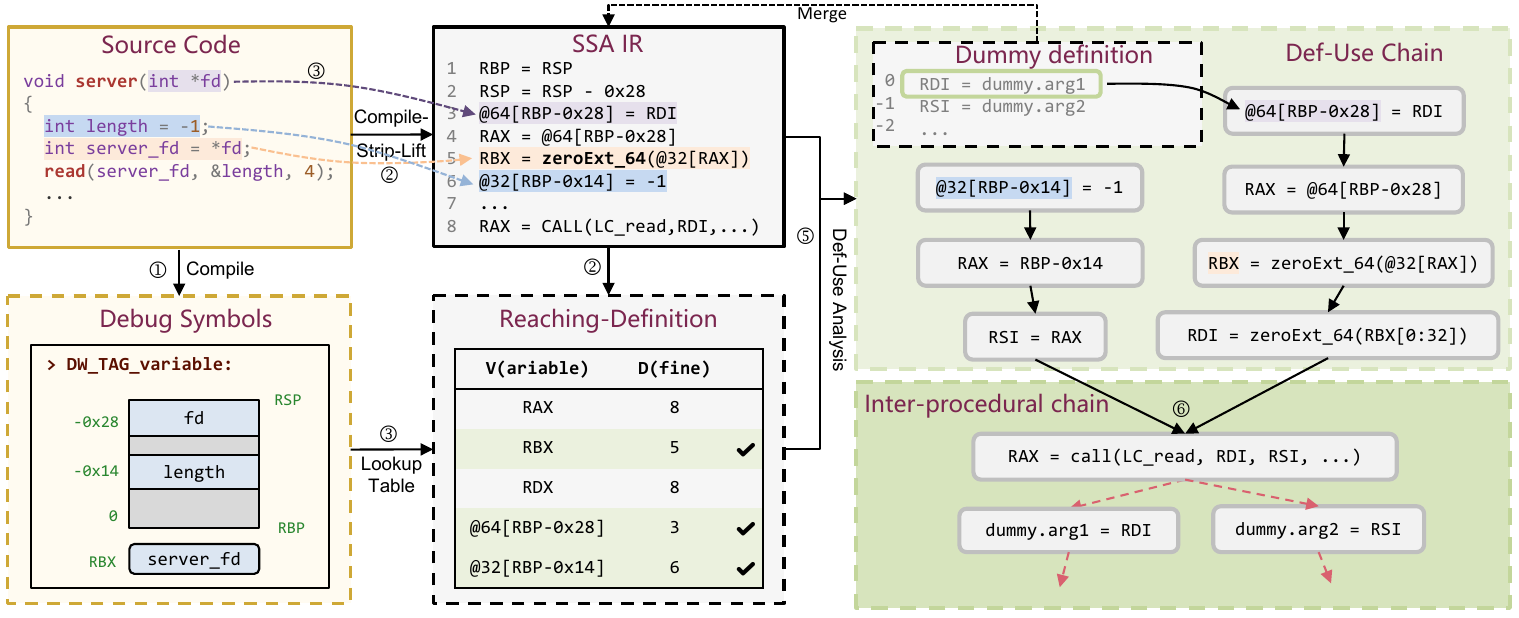}
    \vspace{-1ex}
    \caption{{Overview of Beyond the Edge Program Analysis. Variables in the source code are analyzed by the \textcircled{3} Reaching Definition to maps to SSA IR, and the propagation of the variables is tracked by the \textcircled{5} Def-Use and \textcircled{6} Inter-procedural program analysis. The \colorbox[rgb]{0.925,0.901,0.929}{purple} marks is associated with the variable \texttt{fd}, the \colorbox[rgb]{0.776,0.850,0.945}{blue} markers are associated with the \texttt{Length} variable, and the \colorbox[rgb]{0.992,0.917,0.854}{orange} markers are associated with the variable \texttt{server\_fd}.}}
    \vspace{-2ex}
    \label{fig:overview.BytePA}
\end{figure}

Def-Use analysis \cite{kennedy1979survey} is another static program analysis technique, which is mainly used to trace the relationship between the definition and the use of variables in a program and analyze them along the control flow graph. When the program is executed to point $p$, a Def-Use relationship is established between the variable $R$ read by the statement $s_p$ and the definition $D_R$ that arrives at point $p$. Connecting multiple Def-Use relations constitutes a Def-Use Chain, {as} depicted in Figure \ref{fig:overview.BytePA}, which represents the propagation path of a variable. Def-Use analysis has important application value in compiler optimization, for example, in dead code elimination optimization, by identifying unused variable definitions, redundant code can be effectively detected and removed, thus improving the execution efficiency of the program.

In our scenario, we use the variable definition location obtained from the previous section as the head node, and by implementing Def-Use analysis, we are able to accurately trace the intra-procedural propagation path of the variable, and then construct a complete variable propagation graph. The propagation graph clearly presents the variable's access pattern characteristics and program semantics, and can provide powerful features for the variable type recovery process.

Def-Use analysis allows us to further investigate the actual use of function parameters. In the previous section, we set a dummy definition for all parameters, which provides a maximum list of possible parameters for the function, but in practice, the function will usually have fewer parameters than this maximum. In order to obtain the actual parameters used by a function, we can use Def-Use analysis techniques to track which dummy definitions of parameters are actually used. Those referenced parameters are the real parameters of the function at runtime.

Through the Def-Use program analysis, we have successfully extracted the intra-procedural propagation paths of variables and constructed an accurate variable propagation graph accordingly. In addition, we also deeply analyze the actual parameter list of the function, which provides a solid foundation for inter-procedural analysis.

\subsubsection{\textbf{Inter-procedural Analysis}}
{The objective of our inter-procedural analysis is to establish the variable connection between the caller and the callee, thereby constructing a complete inter-procedural def-use chain. In the SSA$_S$ IR, the argument passing in function calls is explicit. However, this information is completely stripped in the SSA$_M$ IR, where we can only infer that a function call has occurred from a call instruction. We do not know which variables are passed as arguments to the callee or which of these arguments are actually used within the callee. Therefore, the primary task when performing inter-procedural analysis on the SSA$_M$ IR is to first recover the function call context, specifically the argument passing information at the call site.}

We can calculate the address of the target function by using the offset to determine the callee function. For the arguments, the ABI specifies how arguments should be passed in function calls. Therefore, we can recover the maximum possible argument list according to the {calling convention}, and combine it with the actual parameter list of the callee function determined through the Def-Use chain analysis in the previous section, {thereby allowing us to link the caller's arguments to the callee's parameters, and then merge the intra-procedural Def-Use chains to obtain the Variable Propagation Graph.} As shown in Figure \ref{fig:overview.BytePA}, we determine the actual argument list of the function, recover the actual window of the function call, and relay the Def-Use analysis of the arguments between the functions.

According to the target function, we categorize function calls into two types: internal call and external call. An internal call is a call to a function within the same ELF file, which allows access to the function body of the target function for inter-procedural analysis. External call refers to calling functions outside the ELF file, such as library functions, which are not in the same ELF file, and therefore, their function bodies are not available. 

\textbf{Internal Call}: In inter-procedural data flow analysis, we first correlate related functions involved in a function call by destination address and parameter passing. Subsequently, Def-Use analysis is performed within each function to trace the propagation path of each variable. After completing the independent analysis of each function, we merge the variable propagation graphs to form a complete data flow graph. Considering the exponential increase in complexity of the inter-procedural analysis, to ensure the feasibility of the analysis, we set an approximation that only a maximum of two layers of function calls are allowed.

\textbf{External Call}: For the case where the target function is a POSIX API, we can directly infer the type of the arguments by leveraging external knowledge. For example, if a variable is passed as the first argument to the \textit{read} function after dereferencing, it can be determined that the variable is of type \textit{int *}. {These details will be presented in Section \ref{sec:construction.VSG}.} For non-POSIX API function calls, the instructions that prepare the context of the function call can similarly reflect the characteristics of the variable, thus aiding in type recovery.

In addition, the call instruction has {implicit side effects} in program analysis. Prior to the execution of a call instruction, registers, and memory are usually in a particular state. However, the execution of a call instruction triggers a subroutine call, which may contain a large number of instructions that potentially modify the state of registers and memory prior to the call instruction. Therefore, after the call instruction is executed, we cannot simply assume that the state of all registers and memory space remains the same as before the call.

The ABI explicitly specifies the semantics of function calls, stating that the state of the callee-saved registers should remain unchanged before and after the execution of a call instruction, while the state of the caller-saved registers may change. Based on the principle of conservatism in program analysis, in Reaching Definition Analysis, the call instruction is usually regarded as destroying the original definitions of all caller-saved registers while keeping the state of all callee-saved registers unchanged. This treatment ensures the precision and reliability of the program analysis.

In summary, in this section, we carry out an inter-procedural program analysis. First, we recover the target function and the actual parameters of the function calls on the SSA$_M$ IR as a way to correlate the related functions involved in the function calls. Next, we perform a Def-Use analysis of the variables in each function and merge the individual local graphs constructed into a global graph reflecting the propagation properties of the variables across functions. {Although inter-procedural variable propagation is very common and holds significant importance for type recovery, prior work has not involved this key point. However, BytePA achieves this, not only going beyond the edge of a function to perform cross-function analysis, but also beyond the edge of previous work.}

\subsection{Beyond the Edge Type Prediction \label{sec:methodology.byteTR}}

\begin{figure}[!t]
    \centering
    \subfigure[Variable Propagation Graph]{
        \includegraphics[width=0.377\textwidth]{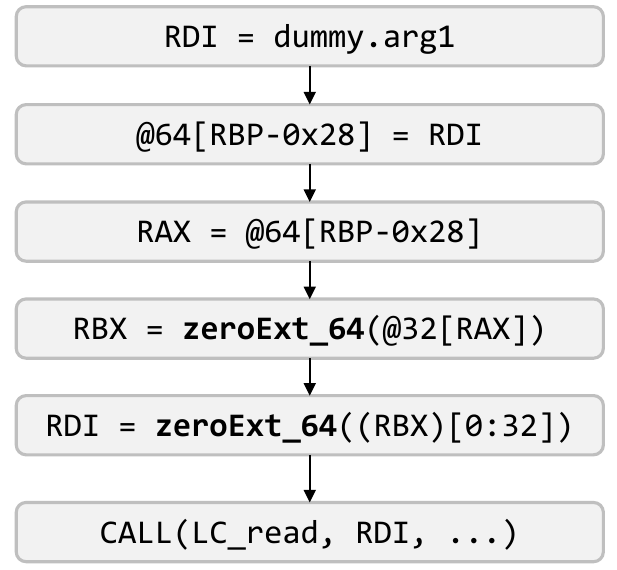}
        \label{fig:variablepropagationgraph}
    }
    \subfigure[Variable Semantic Graph]{
        \includegraphics[width=0.46\textwidth]{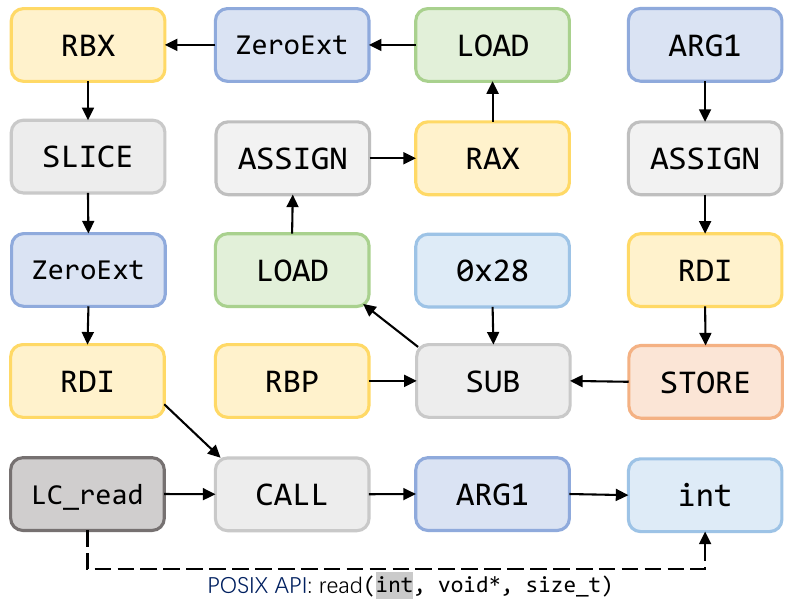}
        \label{fig:variablesemanticgraph}
    }
    \vspace{-1ex}
    \caption{Transformation from Variable Propagation Graph to Variable Semantic Graph}
    \label{fig:variablepropagationgraphandvariablesemanticgraph}
    \vspace{-2ex}
\end{figure}

In the previous section, we derive the inter-procedural variable propagation graphs through program analysis. {Since inter-procedural variables often have long-distance dependencies, traditional graph models can easily suffer from oversmoothing during multi-layer propagation. We choose the {Gated Graph Neural Networks (GGNN)} model because its {Gated Recurrent Unit (GRU)} component can more effectively control information flow during multi-step iteration, helping to retain long-distance information.} In this section, we will present the ByteTP component, which utilizes GGNN to learn the semantic features of variables in order to predict their types.

\subsubsection{\textbf{Construction of Variable Semantic Graph} \label{sec:construction.VSG}}

Recalling the construction of Variable Propagation Graph, we first find the head IR node of the variable definition through Reaching Definition analysis, then use Def-Use program analysis to obtain the propagation path of the variable, and link the propagation path between different functions through inter-procedural analysis, and finally construct the variable propagation graph, as shown in Figure \ref{fig:variablepropagationgraph}. However, since it is difficult to directly perform graph embedding for an individual IR instruction, we need to further unlock the semantics of Variable Propagation Graph to construct the Variable Semantic Graph for type prediction. We argue that the semantics of variable is mainly reflected in the intra-IR structures (e.g., a variable that performs an addition operation with an immediate number usually indicates that it is an integer) as well as in the inter-IR relations instructions (e.g., the Def-Use relationship). By analyzing these features, we can effectively extract the semantic information of variables.

\textbf{Intra-IR Structures}. A single SSA IR such as \texttt{RAX=RBX+1} can be abstracted in its general form as \texttt{lvar=rexpr}. Intuitively, this structure represents the flow of data from the expression on the right to the variable on the left. Among them, \texttt{rexpr} is usually a complex structure of operand variables, which can represent a variety of operations, such as \texttt{ExprOp} for mathematical operations, \texttt{ExprMem} for memory {dereferences}, \texttt{ExprCond} for conditional expressions, and so on. Each \texttt{rexpr} has its own specific parameter structure. For example, \texttt{ExprOp} is parameterized by the variable it operates on, while \texttt{ExprMem} is parameterized by the address expression and the size of the referenced memory.

The core idea of constructing the Intra-IR subgraph is to decompose a complex IR instruction into smaller semantic units, representing their internal data flow with a graph where the key is to connect the rexpr to the lvar node. Specifically, these complex structures are often a combination of various operations, each with a fixed parameter structure. We create a core node with the name of each operation. The parameter nodes are connected to this core node via edges, and the output of the core node is connected to its recursive parent structure. Taking the third example in the Figure \ref{fig:variablepropagationgraph}, \texttt{RAX = @64[RBP-x28]} means loading the 64-bit data at address RBP-0x28 into the register RAX. In this case, the lvar is the register RAX, and the rexpr is a pointer dereference operation. When we convert this to a graph structure, the core of the rexpr is a LOAD node. The input to this node is the memory expression RBP - 0x28, which is recursively expanded into a subgraph represented by a SUB node. Since there are no other operations, the output of the LOAD node is relayed to the target register RAX through an ASSIGN node, as shown in Figure \ref{fig:variablesemanticgraph}. It's worth noting that to clearly illustrate our core method, we have omitted some details, such as the size of the memory load and the order of the edges.

\textbf{Inter-IR Relations}. In variable propagation graph, we maintain inter-IR data flow relations, including the Def-Use relation for variable propagation and the internal call and external call relations for function calls. The Def-Use relation reveals the pattern between the definition of a variable and its use, {which is reflected in the reuse of the lvar of the preceding IR in the rexpr of the succeeding IR. We merge the subgraphs of adjacent IRs using the reused variables, thus completing the data flow connection between adjacent IRs, which then allows us to merge them to obtain the Variable Semantic Graph for a single function.} The function call relation links the data flow between the two functions, {which is a special data flow relation with important features for representing the semantics of a variable. We use a CALL node to connect the caller and callee functions, forming the final inter-procedural Variable Semantic Graph.}

In particular, for external calls, we only keep the call window since the callee function is not available. If the called function is a POSIX API, we set the parameter type corresponding to the API to a solid type node and connect it to the original graph via a directed edge, as shown in Figure \ref{fig:variablesemanticgraph}. This design can provide crucial information for recovering the type of a variable in this case.

\subsubsection{\textbf{Graph Normalization}}

Before feeding the variable semantic graph into the model to learn semantic representations of variables, we need to normalize the nodes and edges for embedding representations. In variable semantic graph, each node is attributed with a token, and we can simply pool all the tokens into a vocabulary and map each label into a learnable embedding vector. However, in practice, there are a large number of different tokens in IR, e.g., various constant numbers, so it is impossible to include all tokens in a vocabulary. We observe that most of the tokens appear only once and are mostly constant values or jump destination addresses, and only a small number of tokens are frequent and indicate the semantics of the program.

To address this problem, we first filter and preserve some high-frequency constant values, such as 0, -1, 1, etc., and uniformly replace the rest of the low-frequency constant values with the special label <CONST>. As for the jump destination address, we replace it with the special label <LOC> directly, as it cannot provide valid semantic information about the variable. In addition, we also exclude tokens that occur only once from the vocabulary. In this way, we not only effectively control the size of the vocabulary, but also are able to better learn the unified semantic representation of constant values and jump addresses.

In variable semantic graph, edges represent paths for data flow. To represent these edges efficiently, we construct a vocabulary of edge types based on their types (e.g., def-use, external call, internal call, etc.). Based on this vocabulary, we map each edge to the corresponding embedding vector for subsequent analysis and processing.

\subsubsection{\textbf{Training GGNN for Type Prediction}}

Graph structures have natural advantages in representing program semantics. First, it can naturally represent data flow through edges, thus clearly showing the data dependencies in a procedure. Second, the symmetry property of graphs allows them to maintain their original topology after inversion, symmetry, or mirroring transformations, thus maintaining the semantics of the program unchanged. In contrast, instruction sequences often lead to changes in program semantics after exchanging two instructions, which lacks the stability of graph structures. Therefore, graph structure has higher robustness and expressive ability in program semantic characterization.

Drawing on relevant research results in the field of graph neural networks \cite{li2019graph, guo2022exploring,he2024code}, we adopt GGNN as the basic framework for node information aggregation. GGNNs efficiently aggregate feature information of neighboring nodes through the mechanism of message passing and dynamically update node states, which feature enables it to well capture the semantic correlation of the data flow between variables. In addition, GGNN uses the Gated Recurrent Unit (GRU) \cite{cho2014learning} as the state update function, which effectively mitigates the long-range dependency problem through its gating mechanism, thus enhancing the effectiveness of inter-procedural analysis.

Given $VSG = G(U, E)$, GGNN maps it to a high-dimensional feature space $D$, as Equation \ref{eq:Methodology.GGNN}, to represent the semantics. Next, we will introduce the GGNN in detail.
\begin{equation}
\begin{aligned}
GGNN: G(U, E) \rightarrow \mathbb{R}^D
\end{aligned}
\label{eq:Methodology.GGNN}
\end{equation}

\textbf{Embedding Initialization}. 
We randomly initialize the initial embedding vectors for nodes and edges, and progressively optimize these vectors during training to learn the representations of each node and edge. Compared to previous work \cite{zhu2024tygr}, which typically relied on manually selected one-hot encoded features (e.g., variable sizes, registers, etc.), our approach is able to capture a more comprehensive variety of semantic information about the variables in learning. Moreover, by combining different training targets, our model can adapt to a variety of downstream tasks, such as variable name inference.

\textbf{Message Passing}. 
The message passing mechanism of GGNN updates the state of a node by exchanging information from neighboring nodes. Specifically, as Equation \ref{eq:Methodology.messagePassing1}, let $x_i$ be the initial vector of the node, whose dimension needs to be less than or equal to the dimension of the output channel, with the shortfall filled with zeros. The ${h}_{i}^{(l)}$ denotes the feature vector of node $i$ at the layer $l$. We compute the message vector ${m}_{i}^{(l+1)}$ of node $i$ at the $l+1$ layer as Equation \ref{eq:Methodology.messagePassing2}, where $j$ is the neighboring node of $i$. Subsequently, this message vector is updated to the feature vector of the next layer by the GRU, as Equation \ref{eq:Methodology.messagePassing3}.
\begin{equation}
\begin{aligned}
\mathbf{h}_i^{(0)} &= \mathbf{x}_i \vert \mathbf{0}, 
\end{aligned}
\label{eq:Methodology.messagePassing1}
\end{equation}
\begin{equation}
\begin{aligned}
\mathbf{m}_{i}^{(l+1)} &= \sum_{j\in\mathcal{N}(i)} e_{j,i} \cdot \boldsymbol{\Theta} \cdot \mathbf{h}_j^{(l)}
\end{aligned}
\label{eq:Methodology.messagePassing2}
\end{equation}
\begin{equation}
\begin{aligned}
\mathbf{h}_{i}^{(l+1)} &= \mathrm{GRU}(\mathbf{m}_{i}^{(l+1)}, \mathbf{h}_{i}^{(l)}).
\end{aligned}
\label{eq:Methodology.messagePassing3}
\end{equation}

\textbf{Aggregation}. 
There are multiple ways of aggregating the features of neighboring nodes in the message passing. In Equation \ref{eq:Methodology.messagePassing2}, we adopt addition as the default method of message aggregation. Different aggregation methods can have a significant impact on the effectiveness of message passing. Apart from addition, commonly used aggregation methods include mean and maximum.

To fully represent the semantic information of the whole graph, we not only need to consider the local feature aggregation between neighboring nodes, but also need to introduce the aggregation mechanism of global node information to generate the final graph embedding representation, as shown in Equation \ref{eq:Methodology.GlobalAggregation}. In this context, the aggregation strategies of the different methods can all be used as key hyperparameters for model tuning.
\begin{equation}
\begin{aligned}
\mathbf{h}_{\text{global}} = \text{Aggregate}\left(\{\mathbf{h}_i^{(L)} \mid i \in \mathcal{U}\}\right)
\end{aligned}
\label{eq:Methodology.GlobalAggregation}
\end{equation}

\textbf{Training}.
Through GGNN, we are able to effectively represent the semantics of variables. In order to adapt to the classification task of type prediction, we add an MLP classification header to the GGNN for mapping the variable feature space to the classification space. During model training, we use a cross-entropy loss function to optimize the model as shown in Equation \ref{eq:Methodology.loss}.

\begin{equation}
\begin{aligned}
L(\mathbf{y}, \hat{\mathbf{y}}) = -\frac{1}{N} \sum_{i=1}^N \sum_{c=1}^C y_{i,c} \log(\hat{y}_{i,c})\end{aligned}
\label{eq:Methodology.loss}
\end{equation}

During the training process, the model calculates the loss values by forward propagation and updates the model parameters by backpropagation, while learning the initial vector representations of the nodes. After training with a large amount of data, the model gradually masters the semantic representations of the variables, and is eventually able to accurately recover variable types.

\subsubsection{{\textbf{Building Graphs for Stripped Binaries}}}
{We previously used debug information to execute the BytePA and ByteTP algorithms, constructing a training dataset. Specifically, debug information provided two crucial pieces of data: (1) the ground truth of the variable graph, and (2) the variable location expressions within the binary code. The ground truth labels were obviously used for model training, while the variable location expressions played a key role in the construction of the VPG. However, when applying ByteTR for type inference on stripped binaries, we lack location expressions compared to the training data construction. It's important to clarify that ByteTR, along with existing type inference works like DIRTY \cite{chen2021augmenting}, StateFormer \cite{pei2021stateformer}, and TYGR \cite{zhu2024tygr}, are not end-to-end systems. Instead, they integrate as plugins or components into decompilers to achieve more accurate type recovery. Therefore, the variable expressions are actually provided by the decompiler, specifically from the variable identification process within the overall decompilation pipeline. From a holistic perspective, variable identification is a prerequisite for type recovery, with its output, the variable's location expression, serving as input for type recovery.}

{In summary, for binaries without debug information, ByteTR first obtains each variable's location expression through the decompiler. The subsequent process is the same as building the training data: it involves using Reaching Definition to determine the location of the variable's definition in the SSA IR, followed by Def-Use analysis to trace the variable's propagation path, and finally, inter-procedural analysis to generate a variable propagation graph. Finally, this variable propagation graph is then transformed and fed into the model to predict the variable's type.}

\section{Evaluation \label{sec:evaluation}}
In this section, we conduct a comprehensive evaluation of ByteTR's performance through a series of experiments aiming to answer the following research questions:
\begin{itemize}
    \item \textbf{RQ1:} How efficient and effective is our method in recovering variable types of binary code? 
    \item \textbf{RQ2:} How does our method compare to baselines in binary code variable type recovery? 
    \item \textbf{RQ3:} How does each core component of our method contribute to overall performance? 
    \item \textbf{RQ4:} How does our method perform in real-world cases? 
\end{itemize}

\subsection{Evaluation Setup \label{sec:evaluation.setup}}
\subsubsection{\textbf{Dataset}\label{sec:evaluation.setup.Dataset}}
We use TYDA \cite{zhu2024tygr} as the experimental dataset. The dataset contains 163,643 binary files with preserved debug symbols, covering four architectures and four optimization options. Compared to previous datasets, TYDA has richer binaries and lower function duplication rates, a quality that has been peer-reviewed. 

It is worth noting that TYDA only contains ELF binary files, so we need to follow up with several preprocessing tasks, including parsing DWARF debugging information, mapping variables to their types, and analyzing them through the BytePA before we can ultimately generate the input data required by the model.

We remove auxiliary functions generated by the compiler, such as \texttt{\_init\_proc}, etc., which cannot be mapped to the source code. We also remove duplicate functions according to binary code hash. In addition, when dividing the training and test sets, we ensure that binary functions from  the same project do not appear in both the training and test sets to avoid data leakage.

\subsubsection{\textbf{Implementation}\label{sec:evaluation.setup.Implementation}}

Our framework is built from about 6,000 lines of Python code, where the inter-procedural program analysis module is implemented based on the miasm framework, while the model training part relies on the PyTorch geometric library. All predefined hyperparameters in the framework are publicly available in our open-source repository.

\subsubsection{\textbf{Environment}\label{sec:evaluation.setup.Environment}}
All experiments in this study were conducted on a server equipped with an Intel(R) Xeon(R) Gold 6330 CPU running the Ubuntu 22.04 operating system. The server is equipped with eight NVIDIA RTX A6000 GPUs for model training.

\subsubsection{\textbf{{Evaluation Settings}}\label{sec:evaluation.setup.settings}}
{To accommodate the differing problem formulations adopted by various type-recovery approaches, We present our evaluation settings to ensure the validity and reproducibility of the experiments. Following prior work, ByteTR is trained on a binary dataset covering four mainstream architectures (x64, x86, ARM, MIPS) and four optimization levels (O0–O3), enabling broad generalization across diverse binary scenarios.
Furthermore, real-world data naturally exhibits substantial distribution imbalance, and directly evaluating models on such raw distributions may yield overly optimistic performance. To mitigate this issue, we balance the number of samples across different categories when evaluating all baselines and ByteTR, ensuring fair and reliable comparisons.
Our evaluation metrics include Precision, Recall, and F1. As type recovery is formulated as a multi-class classification task, we report Macro-F1 for an unbiased comparison across types. Finally, as no existing implementation enables general-purpose LLMs to directly perform variable-type recovery, we design a zero-shot prompting strategy for our experiments. Specifically, we provide each LLM with the assembly instructions and pseudo code of a binary function and instruct it to infer the types of the specified variables.}

\subsection{RQ1: Overall Performance \label{sec:evaluation.RQ1}}

\begin{table}[!t]
\centering
\caption{{Performance comparison across different architectures and optimization levels}}
\label{tab:performance}
\begin{threeparttable}
\begin{tabular}{cccccccc}
\toprule
\multirow{2}{*}{\textbf{ARCH}} & \multirow{2}{*}{\textbf{OPT}} & \multicolumn{3}{c}{\textbf{Effectiveness}} & \multicolumn{3}{c}{\textbf{Efficiency\tnote{1}}} \\
\cmidrule(lr){3-5} \cmidrule(lr){6-8}
& & Precision & Recall & F1-score & Preprocessing & Inference & Latency{\tnote{2}} \\
\midrule
\multirow{4}{*}{x64} 
& O0 & 82.64 & 81.08 & 80.94 & $32.55_{\pm 0.27}$ & $12.46_{\pm 0.19}$ & $45.01_{\pm 0.46}$  \\
& O1 & 80.34 & 79.71 & 77.49 & $24.21_{\pm 0.46}$ & $11.64_{\pm 0.10}$ & $35.85_{\pm 0.56}$  \\
& O2 & 78.95 & 78.37 & 76.85 & $26.96_{\pm 0.36}$ & $ 7.65_{\pm 0.03}$ & $34.61_{\pm 0.39}$  \\
& O3 & 77.86 & 77.75 & 75.97 & $26.42_{\pm 0.36}$ & $11.87_{\pm 0.18}$ & $38.29_{\pm 0.54}$  \\
\midrule
\multirow{4}{*}{x86} 
& O0 & 74.13 & 72.34 & 71.33 & $30.75_{\pm 0.31}$ & $11.53_{\pm 0.05}$ & $42.28_{\pm 0.36}$  \\
& O1 & 73.52 & 74.30 & 70.79 & $21.41_{\pm 0.32}$ & $10.36_{\pm 0.14}$ & $31.77_{\pm 0.46}$  \\
& O2 & 71.75 & 70.26 & 67.36 & $26.63_{\pm 0.05}$ & $ 8.63_{\pm 0.29}$ & $35.26_{\pm 0.34}$  \\
& O3 & 70.93 & 70.49 & 67.81 & $23.86_{\pm 0.08}$ & $ 9.76_{\pm 0.05}$ & $33.62_{\pm 0.13}$  \\
\midrule
\multirow{4}{*}{ARM}
& O0 & \textbf{85.65} & \textbf{84.65} & \textbf{84.26} & $15.45_{\pm 0.41}$ & $11.22_{\pm 0.28}$ & $26.67_{\pm 0.69}$  \\
& O1 & 83.78 & 82.84 & 82.31 & $21.32_{\pm 0.11}$ & $ 9.40_{\pm 0.25}$ & $30.72_{\pm 0.36}$  \\
& O2 & 81.37 & 82.40 & 80.60 & $11.09_{\pm 0.55}$ & $12.03_{\pm 0.10}$ & $23.12_{\pm 0.65}$  \\
& O3 & 79.94 & 81.83 & 79.97 & $14.19_{\pm 0.36}$ & $ 7.48_{\pm 0.09}$ & $21.67_{\pm 0.45}$  \\
\midrule
\multirow{4}{*}{MIPS} 
& O0 & 70.04 & 72.56 & 67.88 & $19.12_{\pm 0.07}$ & $10.27_{\pm 0.12}$ & $29.39_{\pm 0.19}$  \\
& O1 & 68.94 & 70.59 & 68.40 & $16.83_{\pm 0.24}$ & $ 7.48_{\pm 0.20}$ & $24.31_{\pm 0.44}$  \\
& O2 & 68.38 & 67.97 & 64.22 & $15.81_{\pm 0.05}$ & $13.52_{\pm 0.10}$ & $29.33_{\pm 0.16}$  \\
& O3 & 65.19 & 65.20 & 60.89 & $18.35_{\pm 0.12}$ & $11.72_{\pm 0.08}$ & $30.07_{\pm 0.20}$  \\
\bottomrule
\end{tabular}
\begin{tablenotes}
    \item[1] Time unit: milliseconds per function.
    {\item[2] Latency: sum of the time for Processing and Inference.}
\end{tablenotes}
\end{threeparttable}
\vspace{-1ex}
\end{table}

To comprehensively evaluate the performance of ByteTR, we conduct experiments in two dimensions, Effectiveness and Efficiency, and the specific results are detailed in Table \ref{tab:performance}. For the effectiveness evaluation, we focus on the core metrics of precision, recall, and f1-score. For efficiency evaluation, we divided ByteTR's actual running process into two stages: preprocessing {(i.e., program analysis and graph construction)} and model inference. {Specifically, we record the latency of each stage separately and sum them to calculate the total time consumed to infer the type.}

\textbf{Effectiveness Evaluation.} We analyze the effectiveness in two directions: architecture and optimization. ByteTR shows the best performance on the ARM architecture with an average precision of $82.69\%$ and the worst performance on the MIPS architecture with an average precision of only $68.14\%$. As two typical reduced instruction sets, it is interesting to note that they exhibit opposite performance. We observe that the MIPS architecture employs a single memory addressing mode, utilizing base addresses combined with 16-bit offsets, as exemplified by instructions like \texttt{"sw v0, 28(s8)"}. This approach demonstrates a singular methodology in memory management. In contrast, the ARM architecture is much richer in addressing modes and can provide more features to describe the behavior of variables, which are crucial for precisely inferring variable types.  In addition, for x64 and x86, two similar architectures, x64 shows higher precision due to the richer semantics of the instruction set. Recall and f1-score show the same trend as precision, further validating the reliability of the conclusions of our analysis.

From the perspective of optimization, O0 outperforms the other options, which is intuitive since optimization aims to streamline code and reduce memory accesses, making variable type recovery harder. Further observation reveals that the gap between O1, O2, and O3 is smaller than that between O0 and O1, which is consistent with the observation in Figure \ref{fig:pattern.of.storage}. As a whole, the difference in effectiveness between optimization options is smaller than that between architectures, which suggests that we perform better robustness of the optimization options.

\textbf{Efficiency Evaluation.} 
We record the time delays of different stages in terms of function level, this is because the basic unit of program analysis is the function. Calculating the processing time of each variable within a function separately would lead to repeated calculations of function delays, which would not accurately reflect the efficiency of our method.

We find that preprocessing time shows some regularity across architectures and optimization options. In terms of architecture, the preprocessing time for x64 and x86 architectures is on average 10.12 ms higher than ARM and MIPS architectures. Our analysis suggests this is because the x64 and x86 architectures belong to the Complex Instruction Set Architecture (CISC), which leads to a longer time for lifting and program analysis, while the ARM and MIPS architectures belong to the Reduced Instruction Set Architecture (RISC), which is easier to analyze. In terms of optimization options, the O1, O2, and O3 optimization levels have relatively shorter preprocessing times due to the performance optimization of the code.

Inference is timed to approximate the same latency across architectures and optimization options due to the platform independence of the model inference task.

ByteTR uses static program analysis to obtain variable propagation paths and construct semantic graphs, and its overall latency has been optimized to the microsecond level. In contrast, the dynamic micro-execution method employed by Stateformer \cite{pei2021stateformer} has latency on the order of tens of seconds level, while the dynamic execution method of TYGR \cite{zhu2024tygr} has latency in the second level. Thus, ByteTR significantly outperforms both approaches in terms of efficiency.

 \begin{tcolorbox}[colback=gray!5,
                  colframe=black,
                  arc=0.8mm, auto outer arc,
                  boxrule=1pt,
                  boxsep=-2pt
                 ]
\textbf{Answering RQ1:} 
For effectiveness, ByteTR achieves the highest precision of $85.65\%$ at the O0 optimization level of the ARM architecture, while demonstrating excellent robustness for compilation optimization. For efficiency, ByteTR's static program analysis method achieves microsecond processing speeds, significantly outperforming traditional works at second level.

\end{tcolorbox}

\subsection{RQ2: Comparative Evaluation \label{sec:evaluation.RQ2}}
Due to the differences in the way different methods model the binary code type recovery task, we will analyze ByteTR in comparison with baselines one by one.

\subsubsection{\textbf{Compare with DIRTY \cite{chen2021augmenting}}}
DIRTY employs a Transformer-based architecture that contains two core components: pseudo code encoder and data layout encoder for extracting features from binary code. Its unique design enables it to support both variable name recovery and variable type recovery tasks. In evaluating the variable type recovery task, DIRTY uses two subsets of the test set, i.e. function in training and function not in training, as described in their paper \cite{chen2021augmenting}. However, ByteTR does not support function in training prediction.

\begin{table}[t]
\centering
\caption{Precision comparison between ByteTR and DIRTY in x64 Architecture O0 Optimization Option}
\label{tab:compareWithDirty}
\begin{tabular}{l cccccccc}
\toprule
\multirow{3}{*}{\textbf{Method}} & \multicolumn{4}{c}{\textbf{DIRTY's dataset}} & \multicolumn{4}{c}{\textbf{ByteTR's dataset}} \\
\cmidrule(lr){2-5} \cmidrule(lr){6-9}
& \multicolumn{2}{c}{In Train} & \multicolumn{2}{c}{Not In Train} & \multicolumn{2}{c}{In Train} & \multicolumn{2}{c}{Not In Train} \\
& ALL & Struct & ALL & Struct & ALL & Struct & ALL & Struct \\
\midrule
DIRTY & 73.61 & 64.15 & 50.71 & 46.62 & 61.07 & 51.62 & 49.74 & 43.72 \\
\midrule
ByteTR & - & - & \textbf{79.10} & \textbf{71.16} & - & - & \textbf{82.98} & \textbf{76.35} \\
\bottomrule
\end{tabular}
\end{table}

In order to fairly compare the performance of DIRTY and ByteTR, we conduct cross-validation experiments. Specifically, we retrain ByteTR on DIRTY's dataset and reproduce DIRTY on our dataset to ensure a fair comparison. The results of the experiments are shown in Table \ref{tab:compareWithDirty}.

We observe that the DIRTY model performs worse overall on our dataset.
As noted in the study by Pal et al. \cite{pal2024len}, this difference in performance mainly stems from the presence of 65.5\% overlapping samples between the training set and test set of the DIRTY dataset, and this data overlap leads to an overestimation of the model's performance. Experimental results on the DIRTY dataset show that ByteTR outperforms the comparison model by 28.38\% and 24.54\% in predicting ALL data type and Struct type, respectively. In addition, ByteTR performs even better when tested on our dataset, with its precision leading by 33.24\% and 32.63\%, respectively. These results fully demonstrate the significant effectiveness of ByteTR.

\subsubsection{\textbf{Compare with StateFormer \cite{pei2021stateformer}}}

StateFormer implements type recovery by modeling the task as a sequence generation problem and using Generative State Modeling (GSM) pre-trained models to learn the operation semantics of assembly instructions. We compare the effectiveness of ByteTR and StateFormer under four different architectures and four optimization options.

\begin{figure}[!t]
    \centering
    \includegraphics[width=0.88\textwidth]{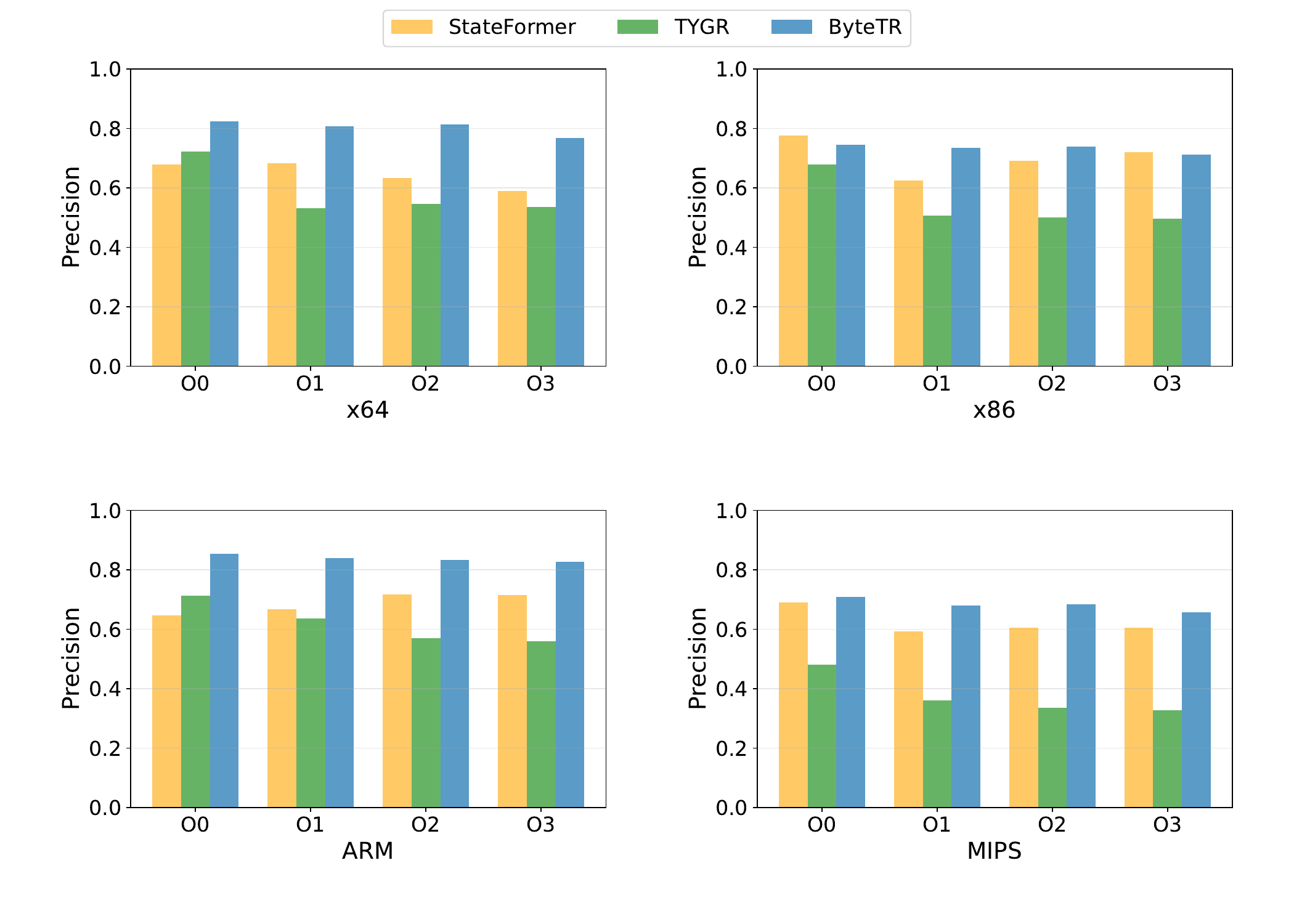}
    \vspace{-2ex}
    \caption{Comparison of precision between StateFormer, TYGR, and ByteTR}
    \vspace{-2ex}
    \label{fig:comparsionwithstateformer}
\end{figure}

As shown in Figure \ref{fig:comparsionwithstateformer}, ByteTR demonstrates a notable advantage in overall performance, with an average precision improvement of $11.93\%$ over StateFormer across all architectures and optimization options. In particular, ByteTR achieves up to $24.80\%$ precision improvement under the O0 optimization configuration for ARM architecture, which significantly outperforms StateFormer. In addition, ByteTR shows strong robustness under different optimization options, whereas StateFormer's performance is much more sensitive and susceptible to compilation optimizations.

\subsubsection{\textbf{Compare with TYGR \cite{zhu2024tygr}}}
TYGR performs symbolic execution analysis of binary code using the angr framework \cite{shoshitaishvili2016sok} to extract memory access patterns of programs. It employs one-hot coding to extract fixed 16 features from each memory access to construct a function-level program access graph and implement type prediction through node classification. However, the fixed feature selection strategy of TYGR may ignore features that are potentially meaningful but not intuitive for type recovery, thus limiting its semantic representation capability. In contrast, our ByteTR approach assigns a label to each token in the variable semantic graph, and is able to capture more implicit semantic information by gradually learning the semantics of each token during the training process. Unlike the node classification method of TYGR, the variable semantic graph embeddings generated by ByteTR are more adaptable and can support a wider range of downstream tasks.

It is worth noting that as TYGR is unable to extract variables stored in registers through compilation optimization (these variables always reside in registers without accessing memory), it is unable to generate the corresponding memory access graph. We choose to manually mark these unprocessable variables as prediction false when calculating the precision.

As shown in Figure \ref{fig:comparsionwithstateformer}, the average precision of ByteTR is 23.29\% higher than that of TYGR. The experimental results show that the performance of TYGR is significantly affected by the processor architecture, and its precision on MIPS architecture is significantly lower than that on ARM architecture. In addition, the inability of TYGR to effectively handle some of the variables optimized by the compiler results in its performance being lower than the unoptimized (O0) case when the optimization options (O1, O2, O3) are enabled. In contrast, ByteTR shows significant advantages in terms of cross-architecture adaptability and compiler optimization robustness.

\subsubsection{{\textbf{Compare with LLMs} \label{sec:comparewithllm}}}
{We compare our method with general LLMs like GPT-4o and Deepseek-V3 for the task of variable type recovery. To accommodate the fixed context length of LLMs, we filter for functions with no more than 200 lines of pseudo code and randomly select variables for the models to query. We evaluate their performance on the x64 architecture across four optimization levels: O0, O1, O2, and O3, using Precision, Recall, and F1-score as our metrics.}

{The results, presented in Table \ref{tab:performance-llm-decompiler}, show that even with pseudo code that provides initial type inference, the LLMs' capability for variable type recovery remains limited. While they could correct obvious type errors in the pseudo code (e.g., for pointers and structs), their overall performance was significantly inferior to our method. For the O0 compilation option, the maximum precision achieved by GPT-4o and Deepseek-V3 was only 64.93\% and 63.81\%, respectively, 14.39\% and 15.51\% lower than our ByteTR method. This demonstrates that despite providing clear instructions and sufficient information, LLMs still struggle with tasks requiring specialized domain knowledge. This highlights the necessity of our approach, which involves specialized modeling and design for this specific type inference challenge.}

\begin{table}[t]
\centering
\caption{{Comparison of LLM-based and Static Program Analysis-based Methods with ByteTR.}}
\vspace{-0.5ex}
\label{tab:performance-llm-decompiler}
\setlength{\tabcolsep}{1mm}
\scalebox{0.86}{
\begin{threeparttable}
\begin{tabular}{cccccccccccccc}
\toprule
\multicolumn{2}{c}{\multirow{2}{*}{\textbf{Method}}} & \multicolumn{3}{c}{\textbf{x64-O0}} & \multicolumn{3}{c}{\textbf{x64-O1}} & \multicolumn{3}{c}{\textbf{x64-O2}} & \multicolumn{3}{c}{\textbf{x64-O3}}\\
\cmidrule(lr){3-5} \cmidrule(lr){6-8} \cmidrule(lr){9-11} \cmidrule(lr){12-14}
& & Pre. & Rec. & F1 & Pre. & Rec. & F1 & Pre. & Rec. & F1 & Pre. & Rec. & F1 \\
\midrule
\multirow{2}{*}{\makecell{LLM}} 
& DeepSeek-V3 & 63.81 & 56.55 & 49.67 & 62.25 & 53.43 & 47.49 & 60.72 & 55.35 & 48.34 & 60.20 & 55.27 & 48.22  \\
& GPT-4o      & 64.93 & 55.10 & 50.65 & 63.87 & 53.93 & 50.01 & 62.79 & 56.71 & 50.59 & 60.59 & 56.44 & 47.72  \\
\midrule
\multirow{2}{*}{\makecell{SPA}\tnote{1}} 
& Ghidra         & 52.25 & 47.18 & 39.98 & 50.29 & 45.16 & 37.54 & 49.46 & 44.20 & 35.93 & 48.75 & 43.87 & 35.90  \\
& IDA Pro   & 55.46 & 48.17 & 43.93 & 53.94 & 48.90 & 42.09 & 52.36 & 47.15 & 40.20 & 51.76 & 49.26 & 41.08  \\
\midrule
\multicolumn{2}{c}{\textbf{ByteTR}} & \textbf{79.32} & \textbf{78.09} & \textbf{75.17} & \textbf{77.66} & \textbf{76.56} & \textbf{74.83} & \textbf{76.23} & \textbf{74.82} & \textbf{71.94} & \textbf{74.83} & \textbf{72.06} & \textbf{69.56}  \\
\bottomrule
\end{tabular}
\begin{tablenotes}
    \item[1] SPA: Static Program Analysis
\end{tablenotes}

\end{threeparttable} }
\vspace{-1ex}
\end{table}

\subsubsection{{\textbf{Compare with Decompilers.}}}
{We select the leading decompilers IDA Pro and Ghidra as our baselines, as they represent the current state-of-the-art in type recovery based on pure program analysis techniques. We evaluate their performance using the same dataset as in section \ref{sec:comparewithllm} and measure their Precision, Recall, and F1-score, for comparison. As shown in Table \ref{tab:performance-llm-decompiler}, both Ghidra and IDA Pro exhibit limited type recovery capabilities. Under the x64-O0 setting, their highest Precision scores were 52.25\% and 55.46\%, respectively, and their performance gradually declined as the compilation optimization level increased. Specifically, because decompilers do not support inter-procedural type propagation, they often incorrectly identify pointer-type variables as \texttt{\_\_int64}, as these variables are not necessarily dereferenced within the current function. In contrast, our proposed method significantly outperforms both Ghidra and IDA Pro across all compilation optimization levels. On average, our method leads Ghidra by 26.82\%  and IDA Pro by 23.62\% in Precision, which demonstrates its superior variable type recovery ability.}

\begin{tcolorbox}[colback=gray!5,
                  colframe=black,
                  arc=0.8mm, auto outer arc,
                  boxrule=1pt,
                  boxsep=-2pt
                 ]
\textbf{Answering RQ2:} 
ByteTR exhibits state-of-the-art precision, outperforming DIRTY, StateFormer, and TYGR by an average of 32.63\%, 11.93\%, and 23.29\%, {as well as DeepSeek-v3 and GPT-4o by 15.26\% and 13.96\%, and Ghidra and IDA Pro by 23.62\% and 26.82\%}, respectively. Moreover, ByteTR exhibits greater generalizability across different architectures  and higher robustness with various optimization options compared to the baselines.
\end{tcolorbox}

\subsection{RQ3: Ablation Study \label{sec:evaluation.RQ3}}

\begin{figure}[!t]
    \centering
    \subfigure[Precision of different calling depth during training]{
        \includegraphics[width=0.4548\textwidth]{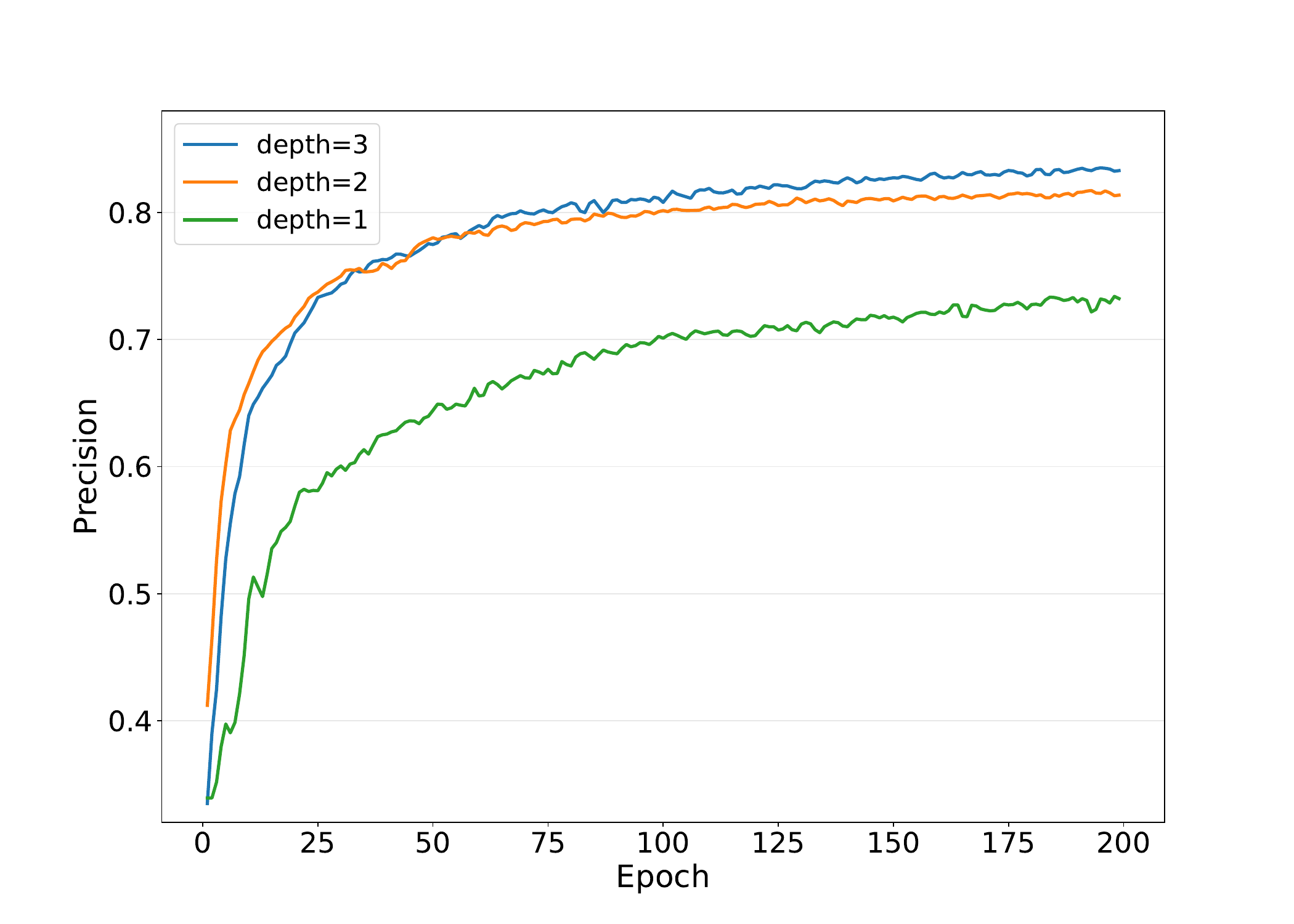}
        \label{fig:ablation1}
    }
    \subfigure[Latency of different calling depth during inference]{
        \includegraphics[width=0.44\textwidth]{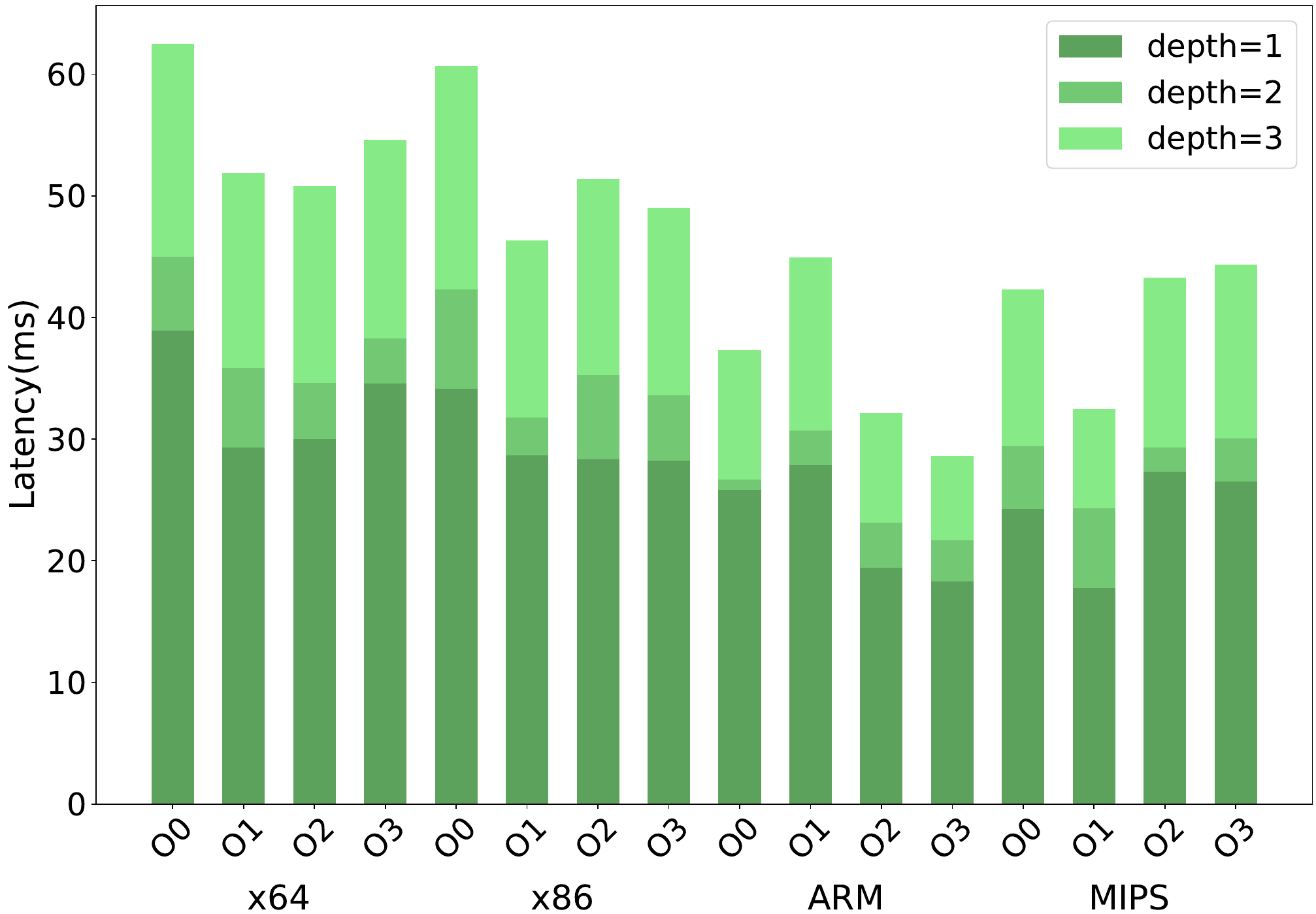}
        \label{fig:efficient1}
    }
    \vspace{-2ex}
    \caption{Precision and latency of different calling depth. The depth=1 means that inter-procedural analysis is not performed. The depth=2 means that only two layers of functions are analyzed, and at 3, three layers of functions are analyzed.}
    \label{fig:ablation}
    \vspace{-3ex}
\end{figure}

In this section, we investigate the impact of inter-procedural analysis on the overall performance of the model through an ablation study. As shown in Figure \ref{fig:ablation}, we analyze in detail the trend of the impact of different function call depths on the model precision and inference latency. It should be noted that Figure \ref{fig:ablation1} shows the training process data when using the O0 optimization level in the ARM architecture.

When the depth of function call is set to 1, it means that it will not perform inter-procedural analysis and only analyze the function in which the variable is defined. whereas the depth settings of 2 and 3 limit the level of call for inter-procedural analysis of variables. As shown in Figure \ref{fig:ablation1}, the precision tends to increase as the depth increases. However, the effect of inter-procedural analysis on the improvement of precision is limited: when the depth of function call is increased from 1 to 2, the precision is improved by $8.46\%$; while when the depth is further increased to 3, it only brings an additional improvement of $2.08\%$. Analyzed from an efficiency perspective, as shown in Figure \ref{fig:efficient1}, the method with a depth of 3 introduces a latency of 13.79ms, which is significantly higher than the latency of 4.53ms with a depth of 2. This trend of latency growth is in line with the theoretical expectation that latency will increase exponentially as the call depth increases.

The results of the study show that inter-procedural analysis has a significant effect on precision. In order to balance effectiveness and efficiency, we set the depth to 2 by default, a configuration that ensures high model performance overall.

\begin{tcolorbox}[colback=gray!5,
                  colframe=black,
                  arc=0.8mm, auto outer arc,
                  boxrule=1pt,
                  boxsep=-2pt
                 ]
\textbf{Answering RQ3:} 
ByteTR performs the inter-procedural analysis and sets the call depth to 2, the precision improves by 8.46\% compared to no inter-procedural analysis, while bringing about an increase in latency of 4.53 ms. However, when the call depth is further increased to 3, the precision improves by only 2.08\%, but the latency increases significantly by 13.73 ms.
\end{tcolorbox}

\begin{figure*}[t]
    \centering
    \subfigure[Source Code]{
        \includegraphics[width=0.46\textwidth]{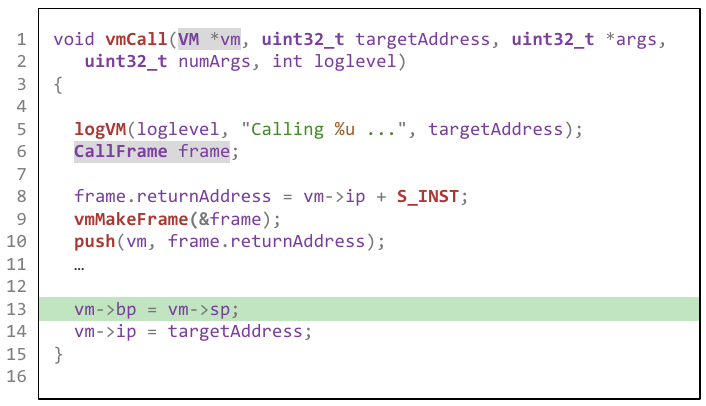}
        \label{fig:readlworldcase1}
    }
    \subfigure[Ghidra Pseudo Code]{
        \includegraphics[width=0.46\textwidth]{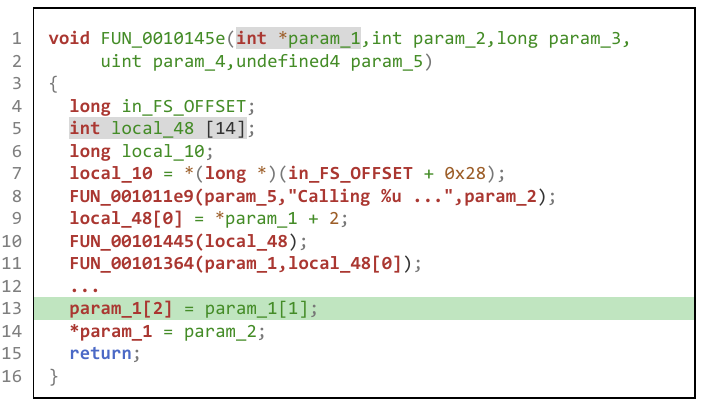}
        \label{fig:realworldcase2}
    }
    \subfigure[IDA Pro Pseudo Code]{
        \includegraphics[width=0.46\textwidth]{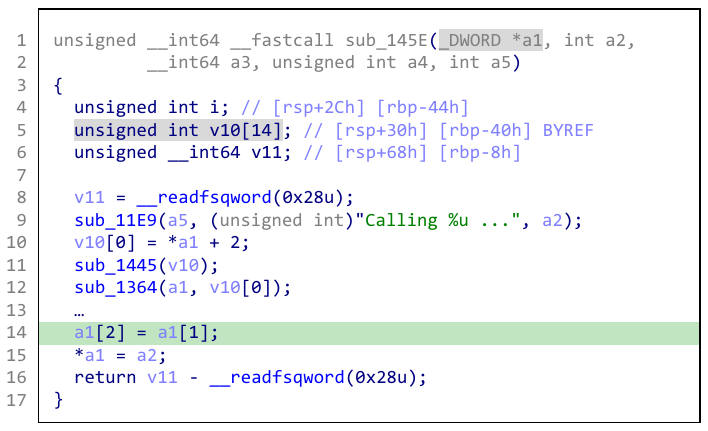}
        \label{fig:realworldcase3}
    }
    \subfigure[{Pseudo Code Optimized by DIRTY}]{
        \includegraphics[width=0.46\textwidth]{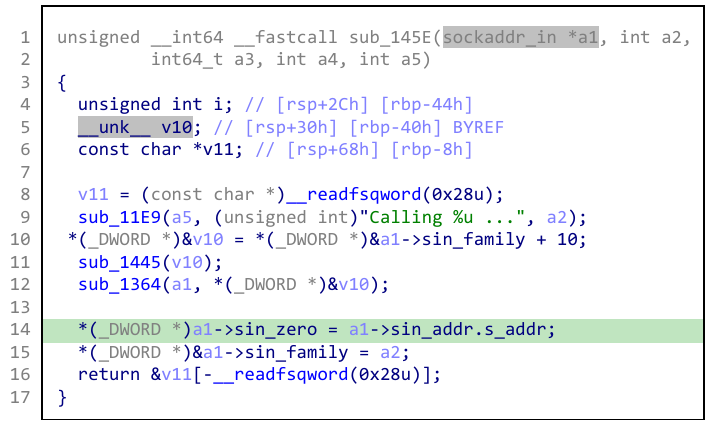}
        \label{fig:realworldcase4}
    }
    \subfigure[{Pseudo Code Optimized by StateFormer}]{
        \includegraphics[width=0.46\textwidth]{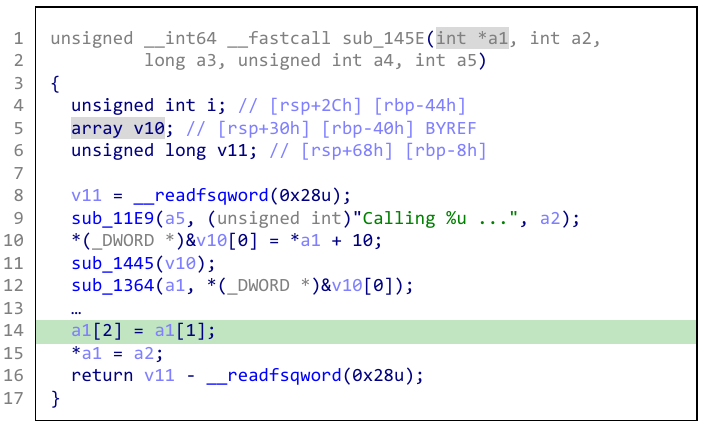}
        \label{fig:realworldcase5}
    }
    \subfigure[Pseudo Code Optimized by ByteTR]{
        \includegraphics[width=0.46\textwidth]{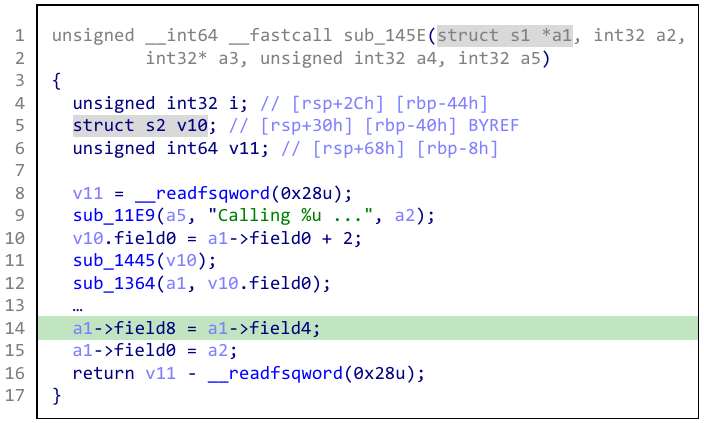}
        \label{fig:realworldcase6}
    }
    \vspace{-2ex}
    \caption{A real-world case in a CTF challenge where a function implements the call instruction of a virtual machine. There is a structure pointer variable \texttt{vm} and a structure variable \texttt{frame} in function \texttt{vmCall}. }
    \vspace{-2ex}
    \label{fig:realworldcase}
\end{figure*}

\subsection{RQ4: Real World Cases \label{sec:evaluation.RQ4}} 

In this section, we apply ByteTR to a vulnerability analysis scenario to evaluate the practical applicability of our method from a qualitative perspective. We specifically selected cases from real-world CTF challenges because they are typically based on real-world applications and contain actual vulnerabilities. As shown in Figure \ref{fig:readlworldcase1}, this function implements a virtual machine call instruction. Specifically, it first sets up the function's stack frame, subsequently pushes the function's return address onto the stack, and finally updates the virtual machine's base pointer and instruction pointer. Due to space constraints, other processing of the stack frame is omitted in this {function}.

We first use the current leading decompilation tools Ghidra \cite{GHIDRA} and IDA Pro \cite{IDA_PRO} to decompile the target function, and successfully obtain the corresponding pseudo code, as shown in Figure \ref{fig:realworldcase2} and Figure \ref{fig:realworldcase3}, {which serve as a set of baselines. We also include the type recovery techniques from DIRTY and StateFormer as another set of baselines, based on their performance in Section \ref{sec:evaluation.RQ2}, as shown in Figure \ref{fig:realworldcase4} and Figure \ref{fig:realworldcase5}. Next, we analyze the performance of these approaches from the following two aspects: (1) Accuracy of Type Recovery, and (2) Readability of Code Structure.}

{Due to the absence of debugging symbols, both decompilers} fail to accurately recover the type information of the struct pointer variable \texttt{a1} and the struct variable \texttt{v10}, {which are highlighted with a gray background.} This defect leads to the misinterpretation  of code accessing struct  members as operations on array and pointer variables, as shown in lines 13 of Ghidra's pseudo code and lines 14 of IDA's pseudo code, highlighted with a green background. This misjudgment seriously affects the understanding of the pseudo code and increases the difficulty of code analysis. {Furthermore, DIRTY also misidentifies the first parameter, \texttt{a1}, as a \texttt{sockaddr\_in} structure pointer, which is completely unrelated to the ground truth, \texttt{vm}. This error occurs because DIRTY's pre-trained model can only recognize types within its training dataset and struggles to infer out-of-vocabulary types. Consequently, it also fails to resolve the struct type for variable v10, labeling it as \texttt{\_\_unk\_\_} (unknown). StateFormer, meanwhile, incorrectly infers parameter \texttt{a1} as \texttt{int*}. It further misidentifies \texttt{v10} as an array because its analysis is confined to the variable's features within the current function's scope. The structural properties of \texttt{v10}, however, are primarily revealed through its usage patterns after being passed to the callee function vmMakeFrame (\texttt{sub\_1445}).} 

In contrast, as shown in Figure~\ref{fig:realworldcase4}, ByteTR successfully identifies the parameter \texttt{a1} as a structure pointer and accurately infers the type for variable \texttt{v10} by tracking the variable's inter-procedural behavior. This process allows it to aggregate contextual information for \texttt{v10} along the entire call chain, leading to the precise inference of its complex structure type. Although ByteTR does not yet support predicting specific structure members, by combining its output with the memory access offset information provided by IDA Pro, we can simply use \texttt{fieldN} to represent the corresponding member, as shown in line~14, and recursively infer the types of its members. As shown in the code line highlighted in green, only ByteTR is able to restore the source code's dereference pattern. The optimized pseudo code significantly improves code readability and helps reverse engineers better understand the program. 

Surprisingly, the variable \texttt{v11}, which is a canary value and automatically generated by the compiler in order to ensure stack memory safety, does not explicitly appear in the source code, and the corresponding variable type can also be recovered by ByteTR, which fully demonstrates our generalization ability. In addition to its practical efficiency, It is worth noting that ByteTR is designed to be decompiler-independent and can be easily adapted to different decompilers as a plug-in to optimize the output pseudo code. 

\begin{tcolorbox}[colback=gray!5,
                  colframe=black,
                  arc=0.8mm, auto outer arc,
                  boxrule=1pt,
                  boxsep=-2pt
                 ]
\textbf{Answering RQ4:} 
{ByteTR not only precisely recovers structure types and their pointers from real-world binary functions, outperforming leading decompilation tools (Ghidra, IDA) and type inference techniques (DIRTY, StateFormer), but its optimized pseudo code also effectively restores variable access patterns, demonstrating significant practical utility.}
\end{tcolorbox}

\section{Discussion and Limitations \label{sec:disandlimi}}

\subsection{Discussion \label{sec:discussion}} 

{This section further discusses ByteTR's performance, showcasing its practical application in recovering composite types, analyzing the effectiveness of Graph Neural Networks, discussing its limitations through two representative failure cases, examining the design principles that ensure its scalability, and explaining other variable-level tasks.}

\subsubsection{{\textbf{Recovering Composite Types with ByteTR}}}
\begin{figure}[!t]
    \centering
    \includegraphics[width=0.95 \textwidth]{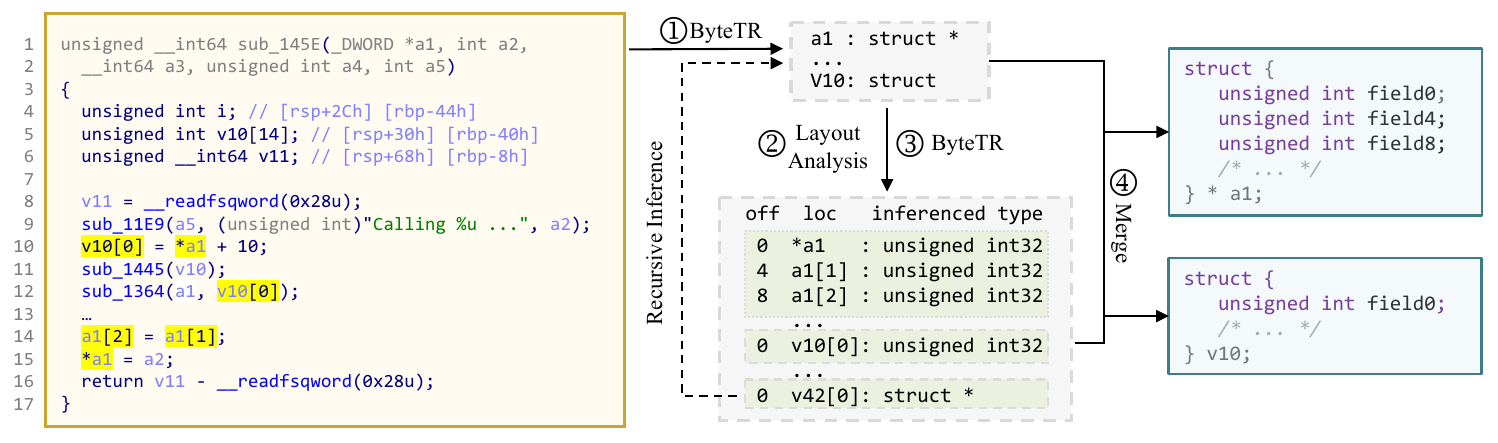}
    \vspace{-2ex}
    \caption{{Recursive Type Inference for Composite Types with ByteTR.}}
    \vspace{-2ex}
    \label{fig:recursiveinference}
\end{figure}

{Even though ByteTR can only determine whether a variable is a composite type and not recover its full definition, by combining it with layout analysis algorithms, we are able to gradually reconstruct the structure of composite types. In this work, we model the recovery of composite types as a “\textbf{Recursive Type Inference}” process. We have already demonstrated our approach for struct type variables in Section \ref{sec:evaluation.RQ4}. Specifically, the process consists of four steps, as illustrated in Figure \ref{fig:recursiveinference}.}

{(1) Infer Variable Type: Use ByteTR to infer the variable's type. If it is a struct or struct pointer, proceed to (2); otherwise, terminate.}

{(2) Perform Layout Analysis: Obtain each member's offset (e.g., 4) and expression (e.g., a1[1]) for the variable.}

{(3) Infer Member Type: Use ByteTR to infer the type of each member. If the result is a struct or struct pointer, go back to (1).}

{(4) Generate Struct Definition: Use field\{offset\} to denote the member at a given offset and merge all members to generate the final struct definition.}

{The resulting struct can be completely patched back into the pseudo code because each member's offset is precisely determined through program analysis. Furthermore, our method naturally handles cases where structs are nested within other structs. For other composite types, such as arrays, we can treat them as structs where all members have the same type, requiring only the inference of a single member's type. Unions are special structs that don't adhere to the non-overlapping memory layout of typical structs, but they can still be processed using this entire workflow.}

{However, this does not fully solve the composite type recovery problem. The members we ultimately recover for a struct are incomplete. For instance, the struct variable \texttt{v10} might have a size of \texttt{0x38}, but the members we identify only account for \texttt{0x4} of that size. While ByteTR can perform inter-procedural analysis and extract members used in different functions along a propagation path, many struct variables simply have members that are never referenced throughout their lifetime. This makes it impossible to recover these missing members, which is a fundamental limitation. We call this the “Incomplete Members” problem and consider it the primary focus of our future work. Furthermore, we believe that the ByteTR system proposed in this work will serve as a strong foundation for future composite type recovery efforts.}

\subsubsection{{\textbf{Effectiveness of Graph Neural Networks.}}}
{Graph Neural Networks (GNNs) are naturally suited for binary code analysis, as variable propagation is fundamentally a data flow problem. Our Variable Semantic Graph (VSG) explicitly models key data flow relationships, such as data assignments, def-use chains, and inter-procedural parameter passing, as a graph structure. In contrast, Large Language Models (LLMs) treat code as a flat sequence of tokens and must learn these complex structural semantics implicitly through large-scale pre-training. By representing these relationships explicitly, the VSG significantly reduces the model's learning complexity and, as our experiments show, exhibits superior robustness to compiler optimizations. Furthermore, our graph model is far more efficient, with orders of magnitude fewer parameters than its LLM-based counterparts. The effectiveness of GNNs, such as GGNN, is well-established in tasks like binary code similarity detection \cite{he2024code}, variable name recovery \cite{lacomis2019dire}, and vulnerability detection \cite{liu2025vul}.}

\subsubsection{{\textbf{Analysis of Mispredictions}}}
{While ByteTR demonstrates strong overall performance, a qualitative analysis of its failure modes is illuminating. Such an analysis reveals the model's current limitations and provides clear directions for future research. Accordingly, we have summarized the mispredictions into two representative types, which are detailed in the following discussion.}

{\textit{Error Case 1: Insufficient Inter-procedural Analysis Depth.} Some key variables are only referenced within specific functions. When our inter-procedural analysis lacks the depth to reach these functions, the corresponding variable features cannot be captured, which leads to type misclassification. For example, if a member of a struct pointer variable is not used, we fail to observe the key features of its dereferencing behavior. This may lead to it being misclassified as an \texttt{int64}, as it exhibits characteristics more similar to an unsigned integer. This issue can be mitigated to some extent by sacrificing a certain degree of efficiency to perform a deeper inter-procedural analysis.}

{\textit{Error Case 2: Unused Sign Feature.} When a signed integer does not utilize its sign feature, its usage pattern is highly similar to that of an unsigned integer. This can lead to misclassification, such as an \texttt{int32} being incorrectly predicted as an \texttt{unsigned int32}. This issue is more prevalent in programs that have undergone compiler optimization. While this is an inherent limitation, this type of misclassification does not severely impact the reverse engineering process, as no relevant logic depends on the variable's sign feature.}

\subsubsection{\textbf{Variable-Level Information Inference.}}
Reviewing the research on function-level semantic representations of binary code, it has made significant progress in tasks such as function name recovery \cite{sha2024llasm, zhang2024enhancing,jin2022symlm,nitin2021direct}, function summary generation \cite{ye2023cp,jin2023binary,shang2024far}, and function vulnerability detection \cite{wang2023binvuldet,luo2023vulhawk}. The ByteTR method proposed in this work innovatively performs program slicing constructs more fine-grained variable-level semantic representations that go beyond the edge of function and applies them to type prediction tasks. This fine-grained representation has good generalization and can effectively and accurately support several variable-level analysis tasks such as variable type recovery, variable name recovery, and variable semantic summary generation.

\subsubsection{{\textbf{Methodological Scalability and Practical Application}}}
{In constructing the Variable Semantic Graph (VSG), we label fine-grained semantic units within each SSA IR instruction rather than treating the entire instruction as a single node. A monolithic approach would create an overly coarse representation, obscuring critical data and control dependencies. Our fine-grained method, in contrast, precisely traces variable interactions and value propagation. Furthermore, the VSG effectively performs variable-level slicing, as it only tracks the propagation of a single variable. This design ensures scalability because complexity in large-scale programs arises from an increase in function size, not an explosive growth in individual variable slices. This characteristic naturally mitigates uncontrolled graph expansion. To further manage computational cost, we also set limits on the inter-procedural analysis depth and the maximum node count.}

\subsection{Limitations \label{sec:limitations}} 

{This section outlines the primary limitations of our method, which we view as direct opportunities for future enhancement. The main challenges involve resolving dynamic control flow, like indirect calls, and expanding the scope of our type analysis.}

\subsubsection{{\textbf{Indirect Calls and Function Pointers}}} 
{In C programs, function pointers introduce indirect calls, whose targets cannot be statically determined at compile time and must be resolved at runtime. Accurately resolving such calls represents a limitation of our current work. This is primarily because precise target resolution requires the integration of sophisticated techniques, such as pointer analysis and dynamic analysis, which would introduce substantial performance overhead. Nevertheless, we fully recognize the importance of supporting indirect calls to enhance the completeness and generality of our method. Therefore, incorporating support for indirect calls is one of the key directions for our future work, requires further investigation.}

\subsubsection{{\textbf{Scope of Target Variables and Types.}}}
{The scope of this study is primarily confined to the type recovery of local variables and function parameters, as they constitute the vast majority of variables within program binaries. Presently, ByteTR does not extend its support to global variables. This limitation stems from the fact that collecting features for global variables necessitates global cross-reference analysis to identify all their usage points. Nevertheless, when the reference locations of global variables are manually provided, ByteTR can successfully construct their propagation graphs and perform accurate type inference. We therefore identify the automated type recovery for global variables as a key area for future work, which will involve designing a dedicated algorithm for cross-reference resolution. Beyond the scope of variables, this work models the type recovery task as a classification problem, limiting the output to a fixed, predefined set of types, making the approach incapable of handling typedef aliases. Although this approach reveals low-level physical details like bit-width and signedness but sacrifices high-level semantic abstraction. In contrast, generative models excel at handling semantically rich, variable-length user-defined types due to their strong semantic understanding and generalization abilities. The potential of using large language models to recover type semantics is therefore a promising research direction. }

\section{Related Works\label{sec:relatedworks}}

\subsection{Binary Code Variable Type Recovery}
Variable types intuitively reflect the attributes of a variable, such as bit size, sign attributes, integer or floating-point numbers, etc., so recovering variable types is crucial for reverse engineers to understand a program. Currently, the basic approach to recovering variable types in binary code is based on determining the representation of a variable in binary code. Prior work \cite{chen2021augmenting, xie2024resym} has been focused on recovering variable types on C-Style pseudo code, while other research has performed type recovery on assembly code \cite{chen2020cati,pei2021stateformer,lee2011tie}, IR \cite{he2018debin,zhu2024tygr,noonan2016polymorphic} or binary programs \cite{he2018debin}. Regarding the type recovery methods, the traditional methods can be mainly divided into two categories: static program analysis \cite{noonan2016polymorphic,elwazeer2013scalable,zhang2021osprey} methods and dynamic program analysis methods \cite{lee2011tie,slowinska2011howard,lin2010automatic}. Currently, mainstream commercial decompilation tools \cite{IDA_PRO,GHIDRA,BIN_JA} have achieved significant results in variable type recovery using static analysis methods. However, there is still a certain degree of deviation between the variable types recovered and the declared types in the original source code. As a result, many deep learning-based data-driven approaches have begun to attempt to recover the mappings between variables and source code types in binary code. Debin \cite{he2018debin} recovers variable types by using Conditional Random Fields, while DIRTY \cite{chen2021augmenting}, StateFormer \cite{pei2021stateformer}, and ReSym \cite{xie2024resym} utilizes a model of the Transformer architecture to generate variable types, and TypeMiner \cite{maier2019typeminer}, CATI \cite{chen2020cati}, TYGR \cite{zhu2024tygr} classify predicted variable types by embedding them in binary code. In particular, OSPREY \cite{zhang2021osprey} aims to recover the layout of structures, while ReSym \cite{xie2024resym} builds on this to further recover the member identifiers. 

In this work, we conducted an empirical  study on this task using real-world programs, uncovering several interesting insights such as unbalanced type distributions and Member Reference Locality of structure variables. Based on our findings, we proposed ByteTR, which decouples variable types to determine the target set of types to be recovered, rather than simply predicting type identifiers {like DIRTY \cite{chen2021augmenting}}. Compared to {prior work \cite{zhu2024tygr, pei2021stateformer, chen2021augmenting, he2018debin}, we significantly improve variable features through inter-procedural program analysis. Our approach focuses on variable propagation, not just memory read/write features like TYGR \cite{zhu2024tygr}, which achieves superior performance in type recovery. Our program analysis algorithm is also based on variable expressions, enabling it to handle both stack-based and register-based variables, a notable advancement over TYGR's symbolic execution, which is limited to stack-based variables. In addition, we use graph neural networks to obtain fine-grained, variable-level semantic representations. Unlike TYGR, which uses static one-hot encodings, our method treats node and edge embeddings as learnable parameters that are updated during training. This dynamic approach substantially enhances the expressiveness of the graph network, achieving state-of-the-art results.}

\subsection{AI for Binary Code Analysis}
As binary code is directly executed by the CPU, at the lowest representation level, its lack of source code mapping makes binary code analysis a challenging but important problem. In the field of software security and reverse engineering, binary code analysis plays a crucial role. With the rise of generative large language model techniques in recent years, researchers have begun to explore solving the problem of binary code analysis using deep learning techniques. With the powerful generative capability of large language models, numerous research efforts have been devoted to recovering lost semantic information in binary code, such as debugging information recovery \cite{he2018debin,chua2017neural}, function name recovery \cite{han2021issta,jin2022symlm, kim2023transformer}, and variable name recovery \cite{pal2024len, nitin2021direct, patrick2023xfl}. In addition, there is a large amount of research focusing on areas such as binary code similarity detection \cite{qasem2023binary,luo2023vulhawk,wang2024clap} and third-party library detection \cite{zhu2022bbdetector,li2023libam}. Artificial intelligence approaches have provided a new paradigm for binary code analysis and are driving development in the field.

\section{Conclusion\label{sec:conclusion}}
In this paper, we first conduct a variety of empirical experiments that delve into the properties of variables and their types in binary code. Based on our findings, we model the type recovery problem of binary code as a classification problem with carefully defined target types. Through further empirical analyses, we observe the prevalence of variable propagation across functions, which prompted us to decide to adopt an inter-procedural analysis approach to track the behavior of variables. To this end, we propose ByteTR, a state-of-the-art framework specialized in recovering type information of variables from stripped binary programs. We conduct comprehensive experiments and show that ByteTR achieves an average precision of 75.84\% in recovering variable types, and its precision is as high as 85.65\% with the O0 optimization option for ARM architectures. In addition, we apply ByteTR to {real-world} CTF challenges, and find that pseudo code optimized by ByteTR significantly improves the readability, which greatly facilitates reverse engineers to understand the program semantics.

\begin{acks}
This work was supported in part by the Natural Science Foundation of China (Grant Nos. U20B2047, 62072421, 62002334, 62102386, and 62121002), the Postdoctoral Fellowship Program of CPSF (Grant No. GZC20252180), and the Anhui Provincial Natural Science Foundation (Grant No. 2508085QF213).
\end{acks}

\bibliographystyle{unsrt}
\bibliography{main.bib}

\begin{thebibliography}{10}

\bibitem{liu2020far}
Zhibo Liu and Shuai Wang.
\newblock How far we have come: Testing decompilation correctness of c
  decompilers.
\newblock In {\em Proceedings of the 29th ACM SIGSOFT International Symposium
  on Software Testing and Analysis}, pages 475--487, 2020.

\bibitem{vadayath2022arbiter}
Jayakrishna Vadayath, Moritz Eckert, Kyle Zeng, Nicolaas Weideman,
  Gokulkrishna~Praveen Menon, Yanick Fratantonio, Davide Balzarotti, Adam
  Doup{\'e}, Tiffany Bao, Ruoyu Wang, et~al.
\newblock Arbiter: Bridging the static and dynamic divide in vulnerability
  discovery on binary programs.
\newblock In {\em 31st USENIX Security Symposium (USENIX Security 22)}, pages
  413--430, 2022.

\bibitem{luo2023vulhawk}
Zhenhao Luo, Pengfei Wang, Baosheng Wang, Yong Tang, Wei Xie, Xu~Zhou, Danjun
  Liu, and Kai Lu.
\newblock Vulhawk: Cross-architecture vulnerability detection with
  entropy-based binary code search.
\newblock In {\em NDSS}, 2023.

\bibitem{avllazagaj2021malware}
Erin Avllazagaj, Ziyun Zhu, Leyla Bilge, Davide Balzarotti, and Tudor
  Dumitraș.
\newblock When malware changed its mind: An empirical study of variable program
  behaviors in the real world.
\newblock In {\em 30th USENIX Security Symposium (USENIX Security 21)}, pages
  3487--3504, 2021.

\bibitem{garcia2018lightweight}
Joshua Garcia, Mahmoud Hammad, and Sam Malek.
\newblock Lightweight, obfuscation-resilient detection and family
  identification of android malware.
\newblock {\em ACM Transactions on Software Engineering and Methodology
  (TOSEM)}, 26(3):1--29, 2018.

\bibitem{cao2020benign}
Michael Cao, Sahar Badihi, Khaled Ahmed, Peiyu Xiong, and Julia Rubin.
\newblock On benign features in malware detection.
\newblock In {\em Proceedings of the 35th IEEE/ACM International Conference on
  Automated Software Engineering}, pages 1234--1238, 2020.

\bibitem{sang2024airtaint}
Qian Sang, Yanhao Wang, Yuwei Liu, Xiangkun Jia, Tiffany Bao, and Purui Su.
\newblock Airtaint: Making dynamic taint analysis faster and easier.
\newblock In {\em 2024 IEEE Symposium on Security and Privacy (SP)}, pages
  3998--4014. IEEE, 2024.

\bibitem{wang2023taintmini}
Chao Wang, Ronny Ko, Yue Zhang, Yuqing Yang, and Zhiqiang Lin.
\newblock Taintmini: Detecting flow of sensitive data in mini-programs with
  static taint analysis.
\newblock In {\em 2023 IEEE/ACM 45th International Conference on Software
  Engineering (ICSE)}, pages 932--944. IEEE, 2023.

\bibitem{liang2022pata}
Jie Liang, Mingzhe Wang, Chijin Zhou, Zhiyong Wu, Yu~Jiang, Jianzhong Liu, Zhe
  Liu, and Jiaguang Sun.
\newblock Pata: Fuzzing with path aware taint analysis.
\newblock In {\em 2022 IEEE Symposium on Security and Privacy (SP)}, pages
  1--17. IEEE, 2022.

\bibitem{IDA_PRO}
{Hex-Rays SA}.
\newblock {IDA Pro}.
\newblock \url{https://www.hex-rays.com/products/ida}, 2023.

\bibitem{GHIDRA}
{NationalSecurityAgency}.
\newblock {Ghidra}.
\newblock \url{https://github.com/NationalSecurityAgency/ghidra}, 2023.

\bibitem{BIN_JA}
{Vector 35}.
\newblock {Binary Ninja}.
\newblock \url{https://binary.ninja/}, 2023.

\bibitem{srinivasan2014recovery}
Venkatesh Srinivasan and Thomas Reps.
\newblock Recovery of class hierarchies and composition relationships from
  machine code.
\newblock In {\em International Conference on Compiler Construction}, pages
  61--84. Springer, 2014.

\bibitem{chen2021augmenting}
Qibin Chen, Jeremy Lacomis, Edward~J. Schwartz, Claire {Le~Goues}, Graham
  Neubig, and Bogdan Vasilescu.
\newblock Augmenting decompiler output with learned variable names and types.
\newblock In {\em 31st USENIX Security Symposium}, Boston, MA, August 2022.

\bibitem{zhu2024tygr}
Chang Zhu, Ziyang Li, Anton Xue, Ati~Priya Bajaj, Wil Gibbs, Yibo Liu, Rajeev
  Alur, Tiffany Bao, Hanjun Dai, Adam Doup{\'e}, et~al.
\newblock $\{$TYGR$\}$: Type inference on stripped binaries using graph neural
  networks.
\newblock In {\em 33rd USENIX Security Symposium (USENIX Security 24)}, pages
  4283--4300, 2024.

\bibitem{pei2021stateformer}
Kexin Pei, Jonas Guan, Matthew Broughton, Zhongtian Chen, Songchen Yao, David
  Williams-King, Vikas Ummadisetty, Junfeng Yang, Baishakhi Ray, and Suman
  Jana.
\newblock Stateformer: fine-grained type recovery from binaries using
  generative state modeling.
\newblock In {\em Proceedings of the 29th ACM Joint Meeting on European
  Software Engineering Conference and Symposium on the Foundations of Software
  Engineering}, pages 690--702, 2021.

\bibitem{he2018debin}
Jingxuan He, Pesho Ivanov, Petar Tsankov, Veselin Raychev, and Martin Vechev.
\newblock Debin: Predicting debug information in stripped binaries.
\newblock In {\em Proceedings of the 2018 ACM SIGSAC Conference on Computer and
  Communications Security}, pages 1667--1680, 2018.

\bibitem{zhang2021osprey}
Zhuo Zhang, Yapeng Ye, Wei You, Guanhong Tao, Wen-chuan Lee, Yonghwi Kwon,
  Yousra Aafer, and Xiangyu Zhang.
\newblock Osprey: Recovery of variable and data structure via probabilistic
  analysis for stripped binary.
\newblock In {\em 2021 IEEE Symposium on Security and Privacy (SP)}, pages
  813--832. IEEE, 2021.

\bibitem{xie2024resym}
Danning Xie, Zhuo Zhang, Nan Jiang, Xiangzhe Xu, Lin Tan, and Xiangyu Zhang.
\newblock Resym: Harnessing llms to recover variable and data structure symbols
  from stripped binaries.
\newblock In {\em Proceedings of the 2024 on ACM SIGSAC Conference on Computer
  and Communications Security}, pages 4554--4568, 2024.

\bibitem{sha2024llasm}
Zihan Sha, Hao Wang, Zeyu Gao, Hui Shu, Bolun Zhang, Ziqing Wang, and Chao
  Zhang.
\newblock llasm: Naming functions in binaries by fusing encoder-only and
  decoder-only llms.
\newblock {\em ACM Transactions on Software Engineering and Methodology}, 2024.

\bibitem{li2015gated}
Yujia Li, Daniel Tarlow, Marc Brockschmidt, and Richard Zemel.
\newblock Gated graph sequence neural networks.
\newblock {\em arXiv preprint arXiv:1511.05493}, 2015.

\bibitem{pal2024len}
Kuntal~Kumar Pal, Ati~Priya Bajaj, Pratyay Banerjee, Audrey Dutcher, Mutsumi
  Nakamura, Zion~Leonahenahe Basque, Himanshu Gupta, Saurabh~Arjun Sawant,
  Ujjwala Anantheswaran, Yan Shoshitaishvili, et~al.
\newblock len or index or count, anything but v1”: Predicting variable names
  in decompilation output with transfer learning.
\newblock In {\em 2024 IEEE Symposium on Security and Privacy (SP)}, pages
  152--152. IEEE Computer Society, 2024.

\bibitem{lacomis2019dire}
Jeremy Lacomis, Pengcheng Yin, Edward Schwartz, Miltiadis Allamanis, Claire
  Le~Goues, Graham Neubig, and Bogdan Vasilescu.
\newblock Dire: A neural approach to decompiled identifier naming.
\newblock In {\em 2019 34th IEEE/ACM International Conference on Automated
  Software Engineering (ASE)}, pages 628--639. IEEE, 2019.

\bibitem{OBJDUMP}
{GNU Binutils}.
\newblock {objdump}.
\newblock \url{https://sourceware.org/binutils/docs/binutils/objdump.html},
  2023.

\bibitem{IEEE754}
Ieee standard for floating-point arithmetic, 2019.
\newblock Revision of IEEE Std 754-2008.

\bibitem{yang2023towards}
Shouguo Yang, Zhengzi Xu, Yang Xiao, Zhe Lang, Wei Tang, Yang Liu, Zhiqiang
  Shi, Hong Li, and Limin Sun.
\newblock Towards practical binary code similarity detection: Vulnerability
  verification via patch semantic analysis.
\newblock {\em ACM Transactions on Software Engineering and Methodology},
  32(6):1--29, 2023.

\bibitem{yang2023Asteria-Pro}
Shouguo Yang, Chaopeng Dong, Yang Xiao, Yiran Cheng, Zhiqiang Shi, Zhi Li, and
  Limin Sun.
\newblock Asteria-pro: Enhancing deep learning-based binary code similarity
  detection by incorporating domain knowledge.
\newblock {\em ACM Trans. Softw. Eng. Methodol.}, 33(1), November 2023.

\bibitem{han2021issta}
Han Gao, Shaoyin Cheng, Yinxing Xue, and Weiming Zhang.
\newblock A lightweight framework for function name reassignment based on
  large-scale stripped binaries.
\newblock In {\em Proceedings of the 30th ACM SIGSOFT International Symposium
  on Software Testing and Analysis (ISSTA)}, ISSTA 2021. Association for
  Computing Machinery, 2021.

\bibitem{song2024bin2summary}
Zirui Song, Jiongyi Chen, and Kehuan Zhang.
\newblock Bin2summary: Beyond function name prediction in stripped binaries
  with functionality-specific code embeddings.
\newblock {\em Proc. ACM Softw. Eng.}, 1(FSE), July 2024.

\bibitem{li2021palmtree}
Xuezixiang Li, Yu~Qu, and Heng Yin.
\newblock Palmtree: Learning an assembly language model for instruction
  embedding.
\newblock In {\em Proceedings of the 2021 ACM SIGSAC Conference on Computer and
  Communications Security}, pages 3236--3251, 2021.

\bibitem{wang2024clap}
Hao Wang, Zeyu Gao, Chao Zhang, Zihan Sha, Mingyang Sun, Yuchen Zhou, Wenyu
  Zhu, Wenju Sun, Han Qiu, and Xi~Xiao.
\newblock Clap: Learning transferable binary code representations with natural
  language supervision.
\newblock In {\em Proceedings of the 33rd ACM SIGSOFT International Symposium
  on Software Testing and Analysis}, pages 503--515, 2024.

\bibitem{shang2024binary}
Xiuwei Shang, Li~Hu, Shaoyin Cheng, Guoqiang Chen, Benlong Wu, Weiming Zhang,
  and Nenghai Yu.
\newblock Binary code similarity detection via graph contrastive learning on
  intermediate representations.
\newblock {\em arXiv preprint arXiv:2410.18561}, 2024.

\bibitem{piantadosi2014zipf}
Steven~T Piantadosi.
\newblock Zipf’s word frequency law in natural language: A critical review
  and future directions.
\newblock {\em Psychonomic bulletin \& review}, 21:1112--1130, 2014.

\bibitem{powers1998applications}
David~MW Powers.
\newblock Applications and explanations of zipf’s law.
\newblock In {\em New methods in language processing and computational natural
  language learning}, 1998.

\bibitem{sano2012zipf}
Yukie Sano, Hideki Takayasu, and Misako Takayasu.
\newblock Zipf' s law and heaps's law can predict the size of potential words.
\newblock {\em Progress of Theoretical Physics Supplement}, 194:202--209, 2012.

\bibitem{zhang2008exploring}
Hongyu Zhang.
\newblock Exploring regularity in source code: Software science and zipf's law.
\newblock In {\em 2008 15th Working Conference on Reverse Engineering}, pages
  101--110. IEEE, 2008.

\bibitem{zhang2009discovering}
Hongyu Zhang.
\newblock Discovering power laws in computer programs.
\newblock {\em Information processing \& management}, 45(4):477--483, 2009.

\bibitem{denning2005locality}
Peter~J Denning.
\newblock The locality principle.
\newblock {\em Communications of the ACM}, 48(7):19--24, 2005.

\bibitem{gcc}
{GNU Project}.
\newblock Gcc, the gnu compiler collection.
\newblock \url{https://gcc.gnu.org/}, 2024.
\newblock Accessed: 2024-01-04.

\bibitem{ren2021unleashing}
Xiaolei Ren, Michael Ho, Jiang Ming, Yu~Lei, and Li~Li.
\newblock Unleashing the hidden power of compiler optimization on binary code
  difference: An empirical study.
\newblock In {\em Proceedings of the 42nd ACM SIGPLAN International Conference
  on Programming Language Design and Implementation}, pages 142--157, 2021.

\bibitem{jiang2024bincola}
Shuai Jiang, Cai Fu, Shuai He, Jianqiang Lv, Lansheng Han, and Hong Hu.
\newblock Bincola: Diversity-sensitive contrastive learning for binary code
  similarity detection.
\newblock {\em IEEE Transactions on Software Engineering}, 2024.

\bibitem{amd64-abi}
{Michael Matz and Jan Hubicka and Andreas Jaeger and Mark Mitchell}.
\newblock System {V} application binary interface: {AMD64} architecture
  processor supplement.
\newblock Technical report, x86-64 ABI, 2018.

\bibitem{miasm}
{CEA IT Security}.
\newblock {Miasm}.
\newblock \url{https://github.com/cea-sec/miasm}, 2023.

\bibitem{kennedy1979survey}
Ken Kennedy.
\newblock {\em A survey of data flow analysis techniques}.
\newblock IBM Thomas J. Watson Research Division, 1979.

\bibitem{li2019graph}
Yujia Li, Chenjie Gu, Thomas Dullien, Oriol Vinyals, and Pushmeet Kohli.
\newblock Graph matching networks for learning the similarity of graph
  structured objects.
\newblock In {\em International conference on machine learning}, pages
  3835--3845. PMLR, 2019.

\bibitem{guo2022exploring}
Yixin Guo, Pengcheng Li, Yingwei Luo, Xiaolin Wang, and Zhenlin Wang.
\newblock Exploring gnn based program embedding technologies for binary related
  tasks.
\newblock In {\em Proceedings of the 30th IEEE/ACM International Conference on
  Program Comprehension}, pages 366--377, 2022.

\bibitem{he2024code}
Haojie He, Xingwei Lin, Ziang Weng, Ruijie Zhao, Shuitao Gan, Libo Chen, Yuede
  Ji, Jiashui Wang, and Zhi Xue.
\newblock Code is not natural language: Unlock the power of semantics-oriented
  graph representation for binary code similarity detection.
\newblock In {\em 33rd USENIX Security Symposium (USENIX Security 24),
  PHILADELPHIA, PA}, 2024.

\bibitem{cho2014learning}
Kyunghyun Cho, Bart Van~Merri{\"e}nboer, Caglar Gulcehre, Dzmitry Bahdanau,
  Fethi Bougares, Holger Schwenk, and Yoshua Bengio.
\newblock Learning phrase representations using rnn encoder-decoder for
  statistical machine translation.
\newblock {\em arXiv preprint arXiv:1406.1078}, 2014.

\bibitem{shoshitaishvili2016sok}
Yan Shoshitaishvili, Ruoyu Wang, Christopher Salls, Nick Stephens, Mario
  Polino, Andrew Dutcher, John Grosen, Siji Feng, Christophe Hauser,
  Christopher Kruegel, et~al.
\newblock Sok:(state of) the art of war: Offensive techniques in binary
  analysis.
\newblock In {\em 2016 IEEE symposium on security and privacy (SP)}, pages
  138--157. IEEE, 2016.

\bibitem{liu2025vul}
Ruitong Liu, Yanbin Wang, Haitao Xu, Jianguo Sun, Fan Zhang, Peiyue Li, and
  Zhenhao Guo.
\newblock Vul-lmgnns: Fusing language models and online-distilled graph neural
  networks for code vulnerability detection.
\newblock {\em Information Fusion}, 115:102748, 2025.

\bibitem{zhang2024enhancing}
Xiaoling Zhang, Zhengzi Xu, Shouguo Yang, Zhi Li, Zhiqiang Shi, and Limin Sun.
\newblock Enhancing function name prediction using votes-based name
  tokenization and multi-task learning.
\newblock {\em Proceedings of the ACM on Software Engineering},
  1(FSE):1679--1702, 2024.

\bibitem{jin2022symlm}
Xin Jin, Kexin Pei, Jun~Yeon Won, and Zhiqiang Lin.
\newblock Symlm: Predicting function names in stripped binaries via
  context-sensitive execution-aware code embeddings.
\newblock In {\em Proceedings of the 2022 ACM SIGSAC Conference on Computer and
  Communications Security}, pages 1631--1645, 2022.

\bibitem{nitin2021direct}
Vikram Nitin, Anthony Saieva, Baishakhi Ray, and Gail Kaiser.
\newblock Direct: A transformer-based model for decompiled variable name
  recov-ery.
\newblock {\em NLP4Prog 2021}, page~48, 2021.

\bibitem{ye2023cp}
Tong Ye, Lingfei Wu, Tengfei Ma, Xuhong Zhang, Yangkai Du, Peiyu Liu, Shouling
  Ji, and Wenhai Wang.
\newblock Cp-bcs: Binary code summarization guided by control flow graph and
  pseudo code.
\newblock {\em arXiv preprint arXiv:2310.16853}, 2023.

\bibitem{jin2023binary}
Xin Jin, Jonathan Larson, Weiwei Yang, and Zhiqiang Lin.
\newblock Binary code summarization: Benchmarking chatgpt/gpt-4 and other large
  language models.
\newblock {\em arXiv preprint arXiv:2312.09601}, 2023.

\bibitem{shang2024far}
Xiuwei Shang, Shaoyin Cheng, Guoqiang Chen, Yanming Zhang, Li~Hu, Xiao Yu,
  Gangyang Li, Weiming Zhang, and Nenghai Yu.
\newblock How far have we gone in binary code understanding using large
  language models.
\newblock In {\em 2024 IEEE International Conference on Software Maintenance
  and Evolution (ICSME)}, pages 1--12. IEEE, 2024.

\bibitem{wang2023binvuldet}
Yan Wang, Peng Jia, Xi~Peng, Cheng Huang, and Jiayong Liu.
\newblock Binvuldet: Detecting vulnerability in binary program via decompiled
  pseudo code and bilstm-attention.
\newblock {\em Computers \& Security}, 125:103023, 2023.

\bibitem{chen2020cati}
Ligeng Chen, Zhongling He, and Bing Mao.
\newblock Cati: Context-assisted type inference from stripped binaries.
\newblock In {\em 2020 50th Annual IEEE/IFIP International Conference on
  Dependable Systems and Networks (DSN)}, pages 88--98. IEEE, 2020.

\bibitem{lee2011tie}
JongHyup Lee, Thanassis Avgerinos, and David Brumley.
\newblock Tie: Principled reverse engineering of types in binary programs.
\newblock 2011.
\newblock Available at
  \url{https://www.ndss-symposium.org/wp-content/uploads/2017/09/lee.pdf}.

\bibitem{noonan2016polymorphic}
Matt Noonan, Alexey Loginov, and David Cok.
\newblock Polymorphic type inference for machine code.
\newblock In {\em Proceedings of the 37th ACM SIGPLAN Conference on Programming
  Language Design and Implementation}, pages 27--41, 2016.

\bibitem{elwazeer2013scalable}
Khaled ElWazeer, Kapil Anand, Aparna Kotha, Matthew Smithson, and Rajeev Barua.
\newblock Scalable variable and data type detection in a binary rewriter.
\newblock In {\em Proceedings of the 34th ACM SIGPLAN conference on Programming
  language design and implementation}, pages 51--60, 2013.

\bibitem{slowinska2011howard}
Asia Slowinska, Traian Stancescu, and Herbert Bos.
\newblock Howard: A dynamic excavator for reverse engineering data structures.
\newblock In {\em NDSS}, 2011.

\bibitem{lin2010automatic}
Zhiqiang Lin, Xiangyu Zhang, and Dongyan Xu.
\newblock Automatic reverse engineering of data structures from binary
  execution.
\newblock In {\em Proceedings of the 11th Annual Information Security
  Symposium}, pages 1--1, 2010.

\bibitem{maier2019typeminer}
Alwin Maier, Hugo Gascon, Christian Wressnegger, and Konrad Rieck.
\newblock Typeminer: Recovering types in binary programs using machine
  learning.
\newblock In {\em Detection of Intrusions and Malware, and Vulnerability
  Assessment: 16th International Conference, DIMVA 2019, Gothenburg, Sweden,
  June 19--20, 2019, Proceedings 16}, pages 288--308. Springer, 2019.

\bibitem{chua2017neural}
Zheng~Leong Chua, Shiqi Shen, Prateek Saxena, and Zhenkai Liang.
\newblock Neural nets can learn function type signatures from binaries.
\newblock In {\em 26th USENIX Security Symposium (USENIX Security 17)}, pages
  99--116, 2017.

\bibitem{kim2023transformer}
Hyunjin Kim, Jinyeong Bak, Kyunghyun Cho, and Hyungjoon Koo.
\newblock A transformer-based function symbol name inference model from an
  assembly language for binary reversing.
\newblock In {\em Proceedings of the 2023 ACM Asia Conference on Computer and
  Communications Security}, pages 951--965, 2023.

\bibitem{patrick2023xfl}
James Patrick-Evans, Moritz Dannehl, and Johannes Kinder.
\newblock Xfl: Naming functions in binaries with extreme multi-label learning.
\newblock In {\em 2023 IEEE Symposium on Security and Privacy (SP)}, pages
  2375--2390. IEEE, 2023.

\bibitem{qasem2023binary}
Abdullah Qasem, Mourad Debbabi, Bernard Lebel, and Marthe Kassouf.
\newblock Binary function clone search in the presence of code obfuscation and
  optimization over multi-cpu architectures.
\newblock In {\em Proceedings of the 2023 acm asia conference on computer and
  communications security}, pages 443--456, 2023.

\bibitem{zhu2022bbdetector}
Xiaoya Zhu, Junfeng Wang, Zhiyang Fang, Xiaokang Yin, and Shengli Liu.
\newblock Bbdetector: A precise and scalable third-party library detection in
  binary executables with fine-grained function-level features.
\newblock {\em Applied Sciences}, 13(1):413, 2022.

\bibitem{li2023libam}
Siyuan Li, Yongpan Wang, Chaopeng Dong, Shouguo Yang, Hong Li, Hao Sun, Zhe
  Lang, Zuxin Chen, Weijie Wang, Hongsong Zhu, et~al.
\newblock Libam: An area matching framework for detecting third-party libraries
  in binaries.
\newblock {\em ACM Transactions on Software Engineering and Methodology},
  33(2):1--35, 2023.

\end{thebibliography}

\end{document}